\documentclass[journal,twoside,web]{ieeecolor}

\let\labelindent\relax 

\usepackage{enumitem}
\usepackage{generic}
\usepackage{float}
\usepackage{cite}
\usepackage{amsmath,amssymb,amsfonts}
\usepackage{mathtools}
\usepackage{bbm}
\usepackage{bm}
\usepackage{url}
\usepackage{algorithmic}
\usepackage{graphicx}
\usepackage[normalem]{ulem}
\usepackage{textcomp}
\usepackage{comment}

\newtheorem{theorem}{Theorem}

\newtheorem{remark}{Remark}
\newtheorem{lemma}{Lemma}

\newtheorem{definition}{Definition}
\newtheorem{example}{Example}
\newtheorem{illusexample}{Illustration}

\usepackage{marginnote,zref-savepos}
\newcounter{labelnote}
\makeatletter
\let\oldmarginnote\marginnote
\renewcommand*{\marginnote}[1]{%
 \begingroup\strut
  \stepcounter{labelnote}\zsaveposx {marginnote-\thelabelnote}
     \ifnum 0\zposx{marginnote-\thelabelnote}<1900000
      \reversemarginpar
      \oldmarginnote{\color{blue}#1}%
     \else
      \normalmarginpar
      \oldmarginnote{\color{blue}#1}%
     \fi
 \endgroup%
}
\makeatother

\begin{document}

\title{Risk-Aware Stability of Discrete-Time Systems}

\author{Margaret P. Chapman$^\dagger$, \IEEEmembership{Member, IEEE}, and Dionysios S. Kalogerias$^\ddagger$, \IEEEmembership{Member, IEEE}
\thanks{The work of M. P. Chapman was supported in part by the Edward S. Rogers Sr. Department of Electrical and Computer Engineering, University of Toronto, and in part by the Natural Sciences and Engineering Research Council of Canada Discovery Grants Program [RGPIN-2022-04140]. Cette recherche a \'{e}t\'{e} financée par le Conseil de Recherches en Sciences Naturelles et en G\'{e}nie du Canada. \textcolor{black}{The work of D. S. Kalogerias was supported in part by a Microsoft gift.}}
\thanks{$^\dagger$M. P. Chapman is with the Edward S. Rogers Sr. Department of Electrical and Computer Engineering, University of Toronto, Toronto, ON M5S 3G4 Canada (email: mchapman@ece.utoronto.ca).}
\thanks{$^\ddagger$D. S. Kalogerias is with the Department of Electrical Engineering, Yale University, New Haven, CT, 06520 USA (email: dionysis.kalogerias@yale.edu).}
\vspace{-20pt}
}

\maketitle
\pagestyle{empty}
\thispagestyle{empty}
\begin{abstract}
We develop a generalized stability framework for stochastic discrete-time systems, where the generality pertains to the ways in which the distribution of the state energy can be characterized. We use tools from finance and operations research called risk functionals (i.e., risk measures) to facilitate diverse distributional characterizations. 
In contrast, classical stochastic stability notions characterize the state energy on average or in probability, which can obscure 
the variability of stochastic system behavior. 
After drawing connections between various risk-aware stability concepts for nonlinear systems, we specialize to linear systems and derive sufficient conditions for the satisfaction of some risk-aware stability properties. These results pertain to real-valued coherent risk functionals and a mean-conditional-variance functional. 
The results reveal novel noise-to-state stability properties, which assess disturbances in ways that reflect the chosen measure of risk. We illustrate the theory through examples about robustness, parameter choices, and state-feedback controllers. 
\end{abstract}


\section{Introduction}\label{secI}
Stability theory of stochastic systems has a long-standing history with contributions from the late 1950s and early 1960s \cite{bertram1959stability, Kats1960OnTS, Krasovskii1961} and is an especially active research area currently \cite{min2020adaptive, qin2020lyapunov, yao2020new, wang2021improved, li2021stochastic, cui2022finite}.
%
    %
%
The applications of stochastic stability theory are numerous, and we list just a few here: trajectory tracking despite modeling errors \cite[p. 426]{kannan2001handbook}, 
%
%
ensuring that state estimation error stays bounded \cite{reif1999stochastic},
%
%
%
target tracking for partially observable systems \cite{dong2022variational},
%
and establishing the convergence of sampling-based optimization algorithms \cite{kushner2003stochastic, kose2021risk}. 

Stochastic stability theory typically assesses system behavior by evaluating the magnitude $|\cdot|$, e.g., Euclidean norm, of the state $x_t$ over time
in probability or in expectation. 
Evaluating the probability of a harmful event, e.g., $|x_t|$ exceeds a desired threshold, provides one approach to analyze stability. 
To analyze stability from a different perspective, one may evaluate the moments of $|x_t|$. 
%
For example, the uniform exponential $p$-stability property (informally) means that the expectation of $|x_t|^p$ decays exponentially at a rate independent of the initial condition. 
However, instead of merely analyzing $E(|x_t|^p)$, we may wish to analyze additional characteristics of the distribution of the \emph{state energy} $|x_t|^p$ over time. For instance, the state energy may represent parameter estimation error in stochastic gradient descent, state estimation error in a Kalman filter, or trajectory tracking error of a robotic system. Rarer larger realizations of the state energy may arise in practice but may be concealed at the analysis stage by restricting one's attention to the expectation of $|x_t|^p$. There may be incomplete knowledge about the system model, suggesting the importance of analyzing the state energy under distributional ambiguity. Also, the classical $p$-stability property ignores the variability of the state energy with respect to its mean, whereas it may be desirable to stabilize such variability in practice.
%

Ultimately, we wish to design control systems whose energy dynamics enjoy specific desirable distributional characteristics. For instance, we may wish the state energy to decay in expectation conditioned on a given fraction of the worst cases. In other settings, we may wish the variance of the state energy to decay, or we may wish the expected state energy to decay in a distributionally robust sense. 
To design such control systems in the future, we require techniques to analyze state energy distributions from more diverse perspectives.

In the literature, the term stability refers to different properties, for example, uniform boundedness of the state trajectory in probability \cite[(D4), p. 31]{kushner1967stochastic}, 
driving the state to the origin within a given time period (i.e., prescribed-time stabilization) \cite{li2021stochastic}, and input-to-state and noise-to-state stability concepts. Since Sontag's pioneering work in the setting of deterministic systems in the late 1980s \cite{sontag1989smooth}, 
%
%
stochastic input-to-state and noise-to-state stability notions have been developed, e.g., see \cite{krstic1998stabilization, tsinias1999concept, deng2001stabilization, huang2009input, zhao2012stochastic, wu2014stability, yao2020new, wang2021improved}. In particular, deterministic input-to-state stability was extended to stochastic noise-to-state stability by Krsti\'{c}, Deng, and colleagues in the late 1990s and early 2000s \cite{krstic1998stabilization, deng2001stabilization}.
%
%
%
In the current work, we emphasize uniform stability notions of the form 
\begin{equation}\label{keystart}
   \rho(\psi(x_t)) \leq a \lambda^t \psi(\mathbf{x}) + b, \quad t \in \{1,2,\dots\},  
\end{equation}
where $\rho$ is a map from random variables to the extended real line, $\psi$ is a state energy function, $\mathbf{x} \in \mathbb{R}^n$ is an initial condition, and $\lambda \in [0, 1)$, $a \geq 0$, and $b \in \mathbb{R}$ are parameters. If $b$ depends on a disturbance process, 
e.g., $b = \sup_{t \in \{0,1,\dots\}}  E(|w_t|^2) < \infty$, 
then the above stability notion describes a noise-to-state stability property.  

In the current work, we develop a uniform stability framework for the analysis of diverse characteristics of the state energy's distribution by permitting a fairly general map $\rho$ in \eqref{keystart} called a risk functional.
A \emph{risk functional} (i.e., risk measure) is a map from a space of random variables representing costs to the extended real line \cite[p. 261]{shapiro2009lectures}. 
Exponential utility is the classical risk functional in the control systems literature pioneered by Whittle in the 1980s and 1990s \cite{whittle1981, whittle1990risk, whittle1991risk}. Other seminal works about exponential utility include those by Howard and Matheson \cite{howard1972risk} and by Jacobson \cite{jacobson1973optimal} in the early 1970s. The exponential utility $\rho_{\text{e},\theta}(Z) \coloneqq \textstyle \frac{-2}{\theta} \log E(\exp(\frac{-\theta}{2}Z))$ of a nonnegative random variable $Z$ with $\theta \neq 0$ is interpreted as a mean-variance approximation, i.e., if the magnitude of $\theta \textbf{var}(Z)$ is small enough, then $\rho_{\text{e},\theta}(Z) \approx E(Z) - \textstyle \frac{\theta}{4} \textbf{var}(Z)$ \cite[p. 765]{whittle1981}. From a viewpoint based on expected utility theory \cite{von1948theory}, $\rho_{\text{e},\theta}$ assumes that an exponential function represents the user's preferences. 
%
%
Such assumptions need not be appropriate for every application, motivating investigations of additional risk-aware criteria. 

Building on the above contributions, additional types of risk functionals have been developed 
\cite{Artzner, rockafellar2002conditional, shapiro2009lectures, ruszczynski2010risk}. A real-valued \emph{coherent} risk functional satisfies four useful properties: convexity, monotonicity, translation equivariance $\rho(Z + a) = \rho(Z) + a$ if $a \in \mathbb{R}$, and positive homogeneity $\rho(\lambda Z) = \lambda \rho(Z)$ if $\lambda > 0$ \cite{Artzner}, \cite[Ch. 6.3]{shapiro2009lectures}. Such a functional also enjoys a distributionally robust representation \cite[Th. 6.6]{shapiro2009lectures}. A common example is the \emph{conditional value-at-risk} of an integrable random variable $Z$, which represents the expectation of $Z$ in a given fraction of the worst cases \cite[Th. 6.2]{shapiro2009lectures}. 
A \emph{mean-dispersion} risk functional takes the form 
 $  \rho(Z) = E(Z) + \lambda D(Z) $, 
where $\lambda \geq 0$ and $D$ is a measure of dispersion relative to $E(Z)$. Examples of $D$ include variance, standard deviation, and upper-semideviation, i.e., $E(\max\{Z- E(Z),0\})$. 
%
%
A \emph{recursive} risk functional of $Z = Z_0 + Z_1 + Z_2$
takes the form 
 $   \rho(Z) = \rho_0(Z_0+ \rho_1(Z_1 + \rho_2(Z_2)) )$,
where $\rho_i$ is a map between spaces of random variables.\footnote{To introduce a recursive risk functional, we have used a time horizon of length $N = 2$ for clarity. More generally, one can use a finite time horizon of length $N \in \mathbb{N}$, an arbitrary natural number, or one can use an infinite time horizon \cite{ruszczynski2010risk,kose2021risk}.} While the structure of a recursive risk functional $\rho$ facilitates the development of risk-aware algorithms, 
it can be difficult to interpret $\rho$, except in special cases, such as when $\rho$ is the expectation. 
%
%
We will further describe specific risk functionals in Section \ref{nonlin}.
%
%
    %

To meet application-specific needs and preferences about managing uncertainty, it is desirable to have at hand a broad collection of maps that assess different distributional characteristics. This is one motivation for the theory of risk functionals and the pursuit of additional research in the intersection of risk functionals and control systems. 
Research in this intersection has been gaining much momentum in recent years. There are new contributions, for example, in risk-aware mean-field games \cite{moon2016linear,saldi2020approximate}, model predictive control \cite{singh2018framework, sopasakis2019risk}, temporal logic \cite{lindemann2021reactive, safaoui2022risk}, barrier functions \cite{ahmadi2021risk}, and optimal-control-based safety analysis \cite{mpctac2021,chapman2021risk}. We refer the reader to our recent survey article about risk-aware optimal control theory \cite{wang2022risk} for additional literature. We will focus the remainder of the literature review on existing work related to risk-aware stability theory, which has been less studied. 
    %
%
%

\subsection{Literature related to risk-aware stability theory}
First, we will describe connections between the exponential utility functional and stability theory, which have been 
known for the past several decades. Then, we will describe the closest related works to the current work: Singh et al. \cite{singh2018framework}, Sopasakis et al. \cite{sopasakis2019risk}, Kishida \cite{kishida2022risk}, and Tsiamis et al. \cite{tsiamis2020risk}. 

In 1988, Glover and Doyle established connections between the linear time-invariant controller that minimizes the infinite-time exponential utility cost functional and the class of stabilizing controllers that satisfy an $\mathcal{H}_\infty$-norm bound in a linear-quadratic setting \cite{glover1988state}. In 1995, James and Baras developed a solution approach for the robust $\mathcal{H}_\infty$ output feedback nonlinear control problem, which includes an asymptotic stability specification, using techniques from an earlier study of a partially observable exponential utility control problem \cite{james1995robust, james1994risk}. In 1999, Pan and Ba{\c{s}}ar provided conditions that guarantee global probabilistic asymptotic stability for nonlinear continuous-time stochastic systems in the context of a long-term exponential utility cost functional \cite{pan1999backstepping}.
In 2000, Dupuis et al. derived an upper bound for the average output power in terms of the input power and the exponential utility functional, which led to a stochastic small gain theorem \cite{dupuis2000robust}. 

Singh et al. \cite{singh2018framework} and Sopasakis et al. \cite{sopasakis2019risk} both studied risk-averse model predictive control with risk-aware stability properties that are defined using a recursive risk functional; see \cite[Def. V.1]{singh2018framework} and \cite[Def. 4]{sopasakis2019risk}, 
respectively. Singh et al. considered a linear system with multiplicative noise, i.e., $x_{t+1} = A(w_t) x_t + B(w_t) u_t$, where there are finitely many realizations of $w_t$ \cite{singh2018framework}. Sopasakis et al. considered a nonlinear generalization with joint state-input constraints \cite{sopasakis2019risk}. 
Lyapunov stability conditions are provided by \cite[Lemma VI.1]{singh2018framework} and \cite[Lemma 5]{sopasakis2019risk}, respectively.
In particular, the lemma \cite[Lemma VI.1]{singh2018framework}
follows directly from a well-established sufficient condition for a classical exponential stability property \cite[Def. 1, Lemma 1]{bernardini2012stabilizing}, which builds on techniques from \cite{morozan1983stabilization}. 
%
In contrast to the above works, we consider linear models with additive noise, which yields noise-to-state stability properties, and 
we emphasize nonrecursive risk functionals and systems with continuous disturbance spaces. 
%

Kishida developed stability theory for linear systems in the context of a specific coherent risk functional, that is, a distributionally robust conditional value-at-risk (CVaR) functional \cite{kishida2022risk}.
The system is linear with additive noise, and the ambiguity set is the family of disturbance distributions with zero mean and known time-invariant covariance \cite[Eq. (2)]{kishida2022risk}. Sufficient conditions for the robust CVaR stability property include the maximum singular value $\sigma_{\text{max}}(A)$ of the dynamics matrix $A$ being strictly less than one \cite[Lemma 3.2]{kishida2022risk}. 
These efforts have inspired us to investigate 
stability conditions in the context of a broad family of coherent risk functionals for linear systems (Theorem \ref{th2}). 
Our linear system model in Theorem \ref{th2} does not require zero-mean or identically distributed disturbances, and
we circumvent the condition of $\sigma_{\text{max}}(A) < 1$ (instead, our analysis relies on Schur stable matrices). 
We study connections between stability definitions that apply to nonlinear systems (Theorem \ref{th1}), and we develop stability theory in the context of a particular noncoherent risk functional as well (Theorem \ref{th3}).

In a linear-quadratic optimal control setting, Tsiamis et al. proposed a constraint on a conditional variance, which admits a quadratic form \cite{tsiamis2020risk}. We propose a risk functional that is a weighted sum of the mean and a conditional variance (Section \ref{nonlin}, Example \ref{ex4}). Then, we incorporate this functional into our risk-aware stability framework, deriving sufficient conditions for stability (Theorem \ref{th3}). The idea of a mean-conditional-variance risk functional comes from the work \cite{tsiamis2020risk}. However, we require different techniques that are useful for risk-aware stability theory, whereas the focus of \cite{tsiamis2020risk} is risk-aware linear-quadratic optimal control theory.

Overall, in contrast to the above works, we advocate for a generalized risk-aware stability viewpoint, in which one can evaluate the state energy in terms of any risk functional. We see value in this degree of generality to accommodate the potential needs of diverse applications, in which different characterizations of state energy distributions may be useful.

\subsection{Contributions}\label{seccont}
We propose a uniform stability analysis framework that characterizes diverse features of the state energy's distribution by permitting $\rho$ in \eqref{keystart} to be a general risk functional. In contrast, the standard paradigm characterizes the average energy or the probability that the state trajectory is close to the origin. Using a generalized risk-aware (uniform) stability property for nonlinear systems (Definition \ref{def2}), our first contribution is to draw connections between different instances of this property for real-valued coherent risk functionals and other common risk functionals (Theorem \ref{th1}). Then, specializing to linear systems, our second contribution is to derive sufficient conditions under which a risk-aware stability property holds, where the property is defined using a real-valued coherent risk functional (Theorem \ref{th2}). These sufficient conditions 
also provide a distributionally robust risk-neutral stability property using the ambiguity set corresponding to the coherent risk functional of interest. Our final contribution is to derive sufficient conditions for stability in the context of a mean-conditional-variance functional (Theorem \ref{th3}). 
%
The last two theorems uncover novel risk-aware noise-to-state stability properties, which have not been reported or studied in the literature to our knowledge. 

\subsection{Organization, notation, and terminology} 
Section \ref{nonlin} studies risk-aware (uniform) stability properties in the context of various risk functionals and discrete-time nonlinear systems. Section \ref{lin} specializes to linear systems and derives sufficient conditions under which risk-aware stability properties are guaranteed for some classes of risk functionals. 
Section \ref{secV} offers illustrative examples. In particular, we provide a simple risk-aware controller that displays empirically a disturbance attenuation effect in the setting of Theorem \ref{th3} (Illustration \ref{illus3}). 
Lastly, Section \ref{secVI} presents concluding remarks, and the Appendix offers supporting technical details.

$\mathbb{N}$ is the set of natural numbers, $\mathbb{N}_0 \coloneqq \{0\} \cup \mathbb{N}$, $\mathbb{R}$ is the real line, and $\bar{\mathbb{R}} \coloneqq \mathbb{R} \cup \{\infty,-\infty\}$ is the extended real line. Given $n \in \mathbb{N}$, $\mathbb{R}^n$ is $n$-dimensional Euclidean space, and $\bar{\mathbb{R}}^{n}$ is extended $n$-dimensional Euclidean space.
$\mathbb{R}^{n \times n}$ is the set of $n \times n$ real matrices. $\mathcal{S}_n$ is the set of $n \times n$ real symmetric positive semidefinite matrices. If $M \in \mathcal{S}_n$, then $M^{\frac{1}{2}}$ is the matrix square root of $M$, i.e., $(M^{\frac{1}{2}})^\top M^{\frac{1}{2}} = M$.
%
%
%
%
If $H \in \mathcal{S}_n$ and $M \in \mathcal{S}_n$, we define the matrix $H_M$ by
\begin{equation}
    H_M \coloneqq (M^{\frac{1}{2}})^\top H M^{\frac{1}{2}}.
\end{equation}
If $M \in \mathcal{S}_n$, then we denote the smallest and largest eigenvalues of $M$ by $\lambda_{\text{min}}(M)$ and $\lambda_{\text{max}}(M)$, respectively. 
%
%
$\mathcal{S}_n^+$ is the set of $n \times n$ real symmetric positive definite matrices.
$|\cdot|$ is the Euclidean norm on $\mathbb{R}^n$. $|\cdot|_{2}$ is the matrix norm induced by the Euclidean norm, i.e., $|M|_{2} \coloneqq \sup_{{|\bf z|} = 1} |M{\bf z}|$ for any $M \in \mathbb{R}^{n \times n}$. 
$I_n$ is the $n \times n$ identity matrix. $0_n$ is the origin of $\mathbb{R}^n$. $\text{tr}(M)$ is the trace of $M \in \mathbb{R}^{n \times n}$.
%
%

If $\mathbb{M}$ is a metric space, then $\mathcal{B}_{\mathbb{M}}$ is the Borel $\sigma$-algebra on $\mathbb{M}$. Given measurable spaces $(\Omega, \mathcal{F})$ and $(\Omega', \mathcal{F}')$, the notation $f : (\Omega, \mathcal{F}) \rightarrow (\Omega', \mathcal{F}')$ means that the function $f : \Omega \rightarrow \Omega'$ is 
$(\mathcal{F},\mathcal{F}')$-measurable. 
If $\Omega$ and $\Omega'$ are metric spaces and $f : (\Omega, \mathcal{B}_{\Omega}) \rightarrow (\Omega', \mathcal{B}_{\Omega'})$, then $f$ is called Borel-measurable. $(\Omega,\mathcal{F},P)$ is a generic probability space, where $E$ is the associated expectation operator. $\mathcal{L}^q(\Omega,\mathcal{F},P)$ is the associated $\mathcal{L}^q$ space, where $\| \cdot \|_q$ is the associated norm and $q \in [1,\infty]$. We denote $\mathcal{L}^q(\Omega,\mathcal{F},P)$ by $\mathcal{L}^q$ for brevity. With slight abuse of notation (which is standard in this case), we use $f \in \mathcal{L}^q$ to denote a function $f : (\Omega,\mathcal{F}) \rightarrow (\mathbb{R},\mathcal{B}_{\mathbb{R}})$ such that $\|f\|_q < \infty$. $\mathcal{L}^{q*}$ is the dual space of $\mathcal{L}^q$, and $\|\cdot\|_{q*}$ is the corresponding norm. The phrase a.e. means almost everywhere with respect to the probability measure $P$. 
%
$\mathcal{I}_F$ is the indicator function of $F \in \mathcal{F}$. w.r.t. means with respect to.

%

We will use basic measurability and integration theorems, such as those from \cite[Ch. 1.5]{ash1972probability},  
without an explicit reference. 
%
%
\section{Generalized risk-aware stability}\label{nonlin}

Consider a fully observable stochastic discrete-time system
\begin{equation}\label{nonlinsys}
    x_{t+1} = f(x_t,w_t), \quad t = 0,1,\dots,
\end{equation}
where $f : \mathbb{R}^n \times \mathbb{R}^d \rightarrow \mathbb{R}^n$ is a Borel-measurable function, $n$ and $d$ are natural numbers, $(x_0,x_1,\dots)$ is an $\mathbb{R}^n$-valued stochastic process, and $(w_0, w_1,\dots)$ is an $\mathbb{R}^d$-valued independent stochastic process. The processes are defined on an arbitrary probability space $(\Omega,\mathcal{F},P)$. 
We assume that the initial state is fixed at an arbitrary vector $\mathbf{x} \in \mathbb{R}^n$.\footnote{More precisely, we assume that the distribution of $x_0$ is the Dirac measure concentrated at $\mathbf{x}$, where $\mathbf{x}$ can be any vector in $\mathbb{R}^n$.}
%
%
%
%

There are many types of energy-like functions in control theory, e.g., class-$\mathcal{K}$ functions, locally positive definite functions, and decrescent functions \cite[Sec. 5.3.1]{sastry2013nonlinear}. We will find the following energy-like function useful in this work.
\begin{definition}[State energy function $\psi$]\label{defstatenergy}
A state energy function $\psi : \mathbb{R}^n \rightarrow \mathbb{R}$ is Borel-measurable, nonnegative, and satisfies $\psi(x) = 0$ if and only if $x = 0_n$. 
\end{definition}

It is natural to choose $\psi$ so that $\psi(x)$ increases as $|x|$ increases. However, we do not need this condition in the current section. In this section, we will study the relationships between different types of risk-aware stability notions for any $\psi$ satisfying Definition \ref{defstatenergy}. In contrast, in Section \ref{lin}, we will specialize to $\psi(x) = x^\top R x$ with $R \in \mathcal{S}_n^+$. 
%
%
%
%

The next definition is a non-risk-aware (i.e., risk-neutral) uniform stability property. 
Variations of the definition to follow are available from \cite[Def. 1]{morozan1983stabilization}, \cite[Def. 1]{bernardini2012stabilizing}, and \cite[Def. 7.3.8]{kannan2001handbook}, for example.  
%
%
%
%
%

\begin{definition}[Risk-neutral stability]\label{def1}
Let a state energy function $\psi$ and a subset $S \subseteq \mathbb{R}^n$ be given. The system
%
\eqref{nonlinsys} 
is uniformly exponentially stable with an offset with respect to the mean in region $S$ if and only if there exist $\lambda \in [0,1)$, $a \in [0,\infty)$, and $b \in \mathbb{R}$ such that
\begin{equation}
    E(\psi(x_t)) \leq a \lambda^t \psi(\mathbf{x}) + b
\end{equation}
for every time $t \in \mathbb{N}$ and initial condition $\mathbf{x} \in S$. 
For brevity, we often write ``stability with respect to the mean'' instead of ``uniform exponential stability with an offset with respect to the mean in region $S$.''
We refer to $\lambda$ as a rate parameter, $a$ as a scale parameter, and $b$ as an offset parameter.
\end{definition}

Definition \ref{def1} describes a uniform property because $(\lambda, a, b)$ does not depend on $\mathbf{x}$ or $t$. While one can also consider a definition in which $(\lambda, a, b)$ can depend on $\mathbf{x}$ or $t$, we focus on the uniform case in this work.
Definition \ref{def1} describes a local property if $S$ is a strictly proper subset of $\mathbb{R}^n$ and a global property otherwise. If $\psi(x) = |x|^2$ and $b = 0$, then Definition \ref{def1} is uniform exponential mean-square stability. 
If $b = c\sup_{t \in \mathbb{N}_0} E(w_t^\top M w_t) < \infty$ for some $c \in (0,\infty)$ and $M \in \mathcal{S}_n^+$, then Definition \ref{def1} describes a risk-neutral noise-to-state stability property. (Additional risk-neutral noise-to-state stability definitions have been introduced in the literature; see \cite[Def. 4.1]{krstic1998stabilization} and \cite[Def. 2]{wu2014stability} for two examples.) 

As described in the Introduction, we are interested in analyzing the state energy over time, which is a random process $(\psi(x_1), \psi(x_2), \dots)$. If we only consider the average state energy, as in Definition \ref{def1}, then we ignore other characteristics of the distribution of the state energy, including the dispersion with respect to the mean (e.g., variance) and the shape of the upper tail (e.g., the mean in a fraction of the worst cases). Hence, we will develop a \emph{risk-aware} stability concept to facilitate the analysis of a broader range of distributional characteristics of the state energy. 

To generalize Definition \ref{def1}, the key observation is that the expectation is a map from a family of random variables to $\bar{\mathbb{R}}$. Hence, we can consider maps that characterize additional features of the distribution of a random variable. Let $\mathcal{Z}$ be a family of random variables defined on $(\Omega,\mathcal{F},P)$, and let $\rho$ be a map from $\mathcal{Z}$ to $\bar{\mathbb{R}}$.\footnote{We prefer smaller realizations of any $Z \in \mathcal{Z}$. That is, $Z$ represents a random cost rather than a random reward. We assume that $\rho(Z) \in \mathbb{R}\cup\{\infty\}$ for every $Z \in \mathcal{Z}$, and we assume that there exists a $Z \in \mathcal{Z}$ such that $\rho(Z)$ is finite, following convention \cite[p. 261]{shapiro2009lectures}. The term risk functional (i.e., risk measure) invokes these assumptions implicitly.} 
$\rho$ is called a \emph{risk functional} (i.e., risk measure), and several examples will follow. The precise definition of $\mathcal{Z}$ will depend on the risk functional of interest. While Examples \ref{ex1}--\ref{ex3} can be found in \cite{shapiro2009lectures}, they are necessary to present to keep the current work self-contained. 

\begin{example}[Value-at-risk]\label{ex1}
Let $\mathcal{Z}$ be the entire family of random variables on $(\Omega,\mathcal{F},P)$. The value-at-risk of $Z \in \mathcal{Z}$ at level $\alpha \in (0,1)$ is defined by
\begin{equation}\label{myvar}
    \text{VaR}_{\alpha}(Z) \coloneqq \inf\{z \in \mathbb{R} : F_Z(z) \geq 1 - \alpha \}, 
\end{equation}
where $F_Z(z) \coloneqq P(\{Z \leq z\})$ is the distribution function of $Z$. The map $(1-\alpha) \mapsto \text{VaR}_{\alpha}(Z)$ is the generalized inverse of $F_Z$, i.e., the quantile function of $F_Z$ \cite[p. 304]{van2000asymptotic}. 
\end{example}
\begin{example}[Conditional value-at-risk]\label{ex2}
The conditional value-at-risk of $Z \in \mathcal{Z} = \mathcal{L}^1$ at level $\alpha \in (0,1]$ is defined by
\begin{equation}\label{cvardefdef}
    \text{CVaR}_{\alpha}(Z) \coloneqq \underset{s \in \mathbb{R}}{\inf} \Big( s + \textstyle\frac{1}{\alpha} E(\max\{Z-s,0\}) \Big).
\end{equation}
$\text{CVaR}_{1}(Z)$ equals $E(Z)$, and the limit of $\text{CVaR}_{\alpha}(Z)$ as $\alpha \rightarrow 0$ from above equals the essential supremum of $Z$ \cite{shapiro2012}. 
The name conditional value-at-risk comes from the following: If $\alpha \in (0,1)$ and $F_Z$ is continuous at $\text{VaR}_{\alpha}(Z)$, then $\text{CVaR}_{\alpha}(Z) = E(Z | Z \geq \text{VaR}_{\alpha}(Z))$ \cite[Th. 6.2]{shapiro2009lectures}. Synonyms include average value-at-risk and expected shortfall. 
\end{example}

The value-at-risk and the conditional value-at-risk assess risk in terms of quantiles. The following functionals assess risk in terms of dispersions relative to the mean.
\begin{example}[Mean-deviation, mean-upper-semideviation]\label{ex3}
Let $q \in [1,\infty)$, $\beta \in [0,\infty)$, and $Z \in \mathcal{Z} = \mathcal{L}^q$ be given. 
The mean-deviation of $Z$ is defined by
\begin{equation}\label{my6}
    \text{MD}_{q,\beta}(Z) \coloneqq E(Z) + \beta \|Z - E(Z) \|_{q},
\end{equation}
while the mean-upper-semideviation of $Z$ is defined by 
\begin{equation}\label{my7}
  \text{MUS}_{q,\beta}(Z) \coloneqq E(Z) + \beta \|\max\{Z - E(Z),0\} \|_{q}.  
\end{equation}
The second term in \eqref{my7} penalizes realizations of $Z$ above the mean but not realizations of $Z$ below the mean. However, the second term in \eqref{my6} does not distinguish between the two cases.
\end{example}

Conditional value-at-risk and the mean-dispersion functionals of Example \ref{ex3} for particular choices of $q$ and $\beta$ belong to the class of real-valued coherent risk functionals \cite[Ch. 6.3]{shapiro2009lectures}. 
As described previously, such a functional satisfies four desirable properties 
and also admits a dual representation as a distributionally robust expectation (to be presented). Let $\varrho$ be a real-valued coherent risk functional defined on $\mathcal{L}^q$ with $q \in [1,\infty)$. Then, there is a bounded family $\mathcal{A} \subset \mathcal{L}^{q*}$ of densities such that 
\begin{equation}\label{dualrep}
    \varrho(Z) = \sup_{\xi \in \mathcal{A}} E(Z \xi), \quad Z \in \mathcal{L}^q,
\end{equation}
by one direction of \cite[Th. 6.6]{shapiro2009lectures}.\footnote{Moreover, the existence of a dual representation \eqref{dualrep} implies that $\varrho$ is a real-valued coherent risk functional; for technical details and information about additional properties of $\mathcal{A}$, please see \cite[Th. 6.6]{shapiro2009lectures}.} 
In particular, every $\xi \in \mathcal{A}$ satisfies $\xi \in \mathcal{L}^{q*}$, $E(\xi) = 1$, and $\xi \geq 0$ a.e. \cite[Eq. (6.38)]{shapiro2009lectures}. 
%
Using \eqref{dualrep}, one can show that if $Z = 0$ a.e., then $\varrho(Z) = 0$.
We will find the exact forms of $\mathcal{A}$ for Examples \ref{ex2} and \ref{ex3} 
useful for Theorem \ref{th1} to come.
\begin{remark}[Special cases of $\mathcal{A}$]\label{remark1}
For conditional value-at-risk at level $\alpha \in (0,1)$, $\mathcal{A}$ is given by \cite[Eq. (6.70)]{shapiro2009lectures} 
\begin{equation}\label{5}
    \mathcal{A} = \left\{\xi \in \mathcal{L}^{\infty} : 0\leq \xi \leq \textstyle\frac{1}{\alpha} \text{ a.e.}, \; E(\xi) = 1 \right\}.
\end{equation}
Note that $\xi \in  \mathcal{A}$ \eqref{5} implies that $|\xi| \leq \frac{1}{\alpha}$ a.e. Here, the upper bound $\frac{1}{\alpha}$ is hyperbolic in the parameter $\alpha$. In the following two cases, however, we will present an upper bound that is affine in the corresponding parameter.

For $q = 1$ and $\beta \in [0,\frac{1}{2}]$, mean-deviation is real-valued and coherent \cite[Ex. 6.19]{shapiro2009lectures}, and $\mathcal{A}$ is given by \cite[Eq. (6.90)]{shapiro2009lectures} 
\begin{equation}\label{6}
    \mathcal{A} = \left\{\xi \in \mathcal{L}^{\infty} : \xi = 1 + \eta - E(\eta), \; \|\eta\|_{\infty} \leq \beta \right\}.
\end{equation}
If $\xi \in \mathcal{A}$ \eqref{6}, then $0 \leq \xi \leq 1 + 2\beta$ a.e. The upper bound $1 + 2\beta$ is affine in $\beta \in [0,\frac{1}{2}]$. 

For $q \in [1,\infty)$ and $\beta \in [0,1]$, mean-upper-semideviation is real-valued and coherent \cite[Ex. 6.20]{shapiro2009lectures}. The corresponding family $\mathcal{A}$ of densities is given by \cite[Eq. (6.96)]{shapiro2009lectures} 
\begin{equation}\label{7}
    \mathcal{A} = \{\xi \in \mathcal{L}^{q*} : \xi = 1 + \eta - E(\eta), \; \|\eta\|_{q*} \leq \beta, \; \eta \geq 0 \text{ a.e.}\}.
\end{equation}
In particular, if $q = 1$ and $\xi \in \mathcal{A}$ \eqref{7}, then $0 \leq \xi \leq 1 + 2\beta$ a.e. The upper bound is affine in $\beta \in [0,1]$.
\end{remark}

The next risk functional $\rho_\nu$ to be described, which takes inspiration from \cite{tsiamis2020risk}, is related to Example \ref{ex3} because it also assesses risk in terms of a dispersion relative to the mean. 
While $\rho_\nu$ is noncoherent in general, 
it enjoys a special meaning in dynamical systems applications, as it depends on a sub $\sigma$-algebra, which can encode the history of a stochastic process. 
%

\begin{example}[Mean-conditional-variance $\rho_\nu$]\label{ex4}
Given $\nu \in [0,\infty)$ and a sub $\sigma$-algebra $\mathcal{F}_i$ of $\mathcal{F}$, the mean-conditional-variance of $Z \in \mathcal{Z} = \mathcal{L}^1$ is defined by
\begin{equation}
    \rho_{\nu}(Z) \coloneqq E(Z) + \nu E(\Delta_t^2),
\end{equation}
where $\Delta_t$ is a real-valued prediction error defined by
\begin{align}
    \Delta_t & \coloneqq E(Z|\mathcal{F}_i)  \mathcal{I}_{F_i} - Z, \label{Deltat}\\
    F_i & \coloneqq \{ \omega \in \Omega : E(Z|\mathcal{F}_i)(\omega) \in \mathbb{R}\}.
\end{align}
%
%
Note that $\rho_{\nu}(Z)$ can be $\infty$. $E(\Delta_t^2)$ is called a conditional variance because $\Delta_t = E(Z|\mathcal{F}_i)  - Z$ a.e.
%
%
%
%
%
%
\end{example}

Any risk functional, e.g., see Examples \ref{ex1}--\ref{ex4}, can be incorporated into the following definition. 
\begin{definition}[Risk-aware stability]\label{def2}
Let a risk functional $\rho : \mathcal{Z} \rightarrow \bar{\mathbb{R}}$, a state energy function $\psi$, and a subset $S \subseteq \mathbb{R}^n$ be given.
The system \eqref{nonlinsys} is uniformly exponentially stable with an offset with respect to $\rho$ in region $S$ if and only if $\psi(x_t) \in \mathcal{Z}$ for every $t \in \mathbb{N}$ and there exist $\lambda \in [0,1)$, $a \in [0,\infty)$, and $b \in \mathbb{R}$ such that
\begin{equation}
    \rho(\psi(x_t)) \leq a \lambda^t \psi(\mathbf{x}) + b
\end{equation}
for every time $t \in \mathbb{N}$ and initial condition $\mathbf{x} \in S$. 
For brevity, we often write ``stability with respect to $\rho$'' instead of ``uniform exponential stability with an offset with respect to $\rho$ in region $S$.'' As before, we refer to $\lambda$ as a rate parameter, $a$ as a scale parameter, and $b$ as an offset parameter.
\end{definition}

In Section \ref{lin}, we will see how $b$ can depend on the disturbance process and the chosen measure of risk, leading to a risk-aware noise-to-state stability property.\footnote{In this work, we do not need to write Definition \ref{def2} in terms of the existence of class-$\mathcal{K}$ and class-$\mathcal{K}\mathcal{L}$ functions. In future work, using these function classes will likely be useful for extending the results of Section \ref{lin} (Theorem \ref{th2} and Theorem \ref{th3}) to nonlinear systems.} 
%
In Definition \ref{def2}, the condition $\psi(x_t) \in \mathcal{Z}$ for every $t \in \mathbb{N}$ is required because the domain of $\rho$ need not be the entire family of random variables on $(\Omega,\mathcal{F},P)$, e.g., see Examples \ref{ex2}--\ref{ex4}. In the special case of $\rho$ being the expectation and $\mathcal{Z}$ being the entire family of random variables on $(\Omega,\mathcal{F},P)$, Definition \ref{def2} reduces to Definition \ref{def1}. 
%
%
While Definition \ref{def2} is inspired by \cite[Def. V.1]{singh2018framework}, \cite[Def. 4]{sopasakis2019risk}, and \cite[Def. 3.1]{kishida2022risk}, Definition \ref{def2} invokes a generalized viewpoint because $\rho$ can be any risk functional. 
In contrast, a recursive risk functional was considered in \cite[Def. V.1]{singh2018framework} and \cite[Def. 4]{sopasakis2019risk}, and a distributionally robust CVaR functional was considered in \cite[Def. 3.1]{kishida2022risk}. 
%
%
The next theorem will demonstrate connections between several instances of Definition \ref{def2}.


\begin{theorem}[Analysis of Definition \ref{def2}]\label{th1}
Consider the discrete-time nonlinear system \eqref{nonlinsys}. Let a state energy function $\psi$, a subset $S \subseteq \mathbb{R}^n$, scalars $\alpha \in (0,1)$ and $q \in [1,\infty)$, and a real-valued coherent risk functional $\varrho$ on $\mathcal{L}^q$ be given. Then, the following statements hold:
\begin{enumerate}
    \item Stability w.r.t. $\text{CVaR}_{\alpha}$ in region $S$ with parameters $(\lambda,a,b)$ implies the probabilistic stability property
  $  P(\{\psi(x_t) \leq a \lambda^t \psi(\mathbf{x}) + b\}) \geq 1-\alpha $ 
for every $t \in \mathbb{N}$ and $\mathbf{x} \in S$.
\item Stability w.r.t. the mean with parameters $(\lambda,a,b)$ implies stability w.r.t. $\text{CVaR}_{\alpha}$ with parameters $(\lambda,\frac{a}{\alpha},\frac{b}{\alpha})$.
%
\item Stability w.r.t. the mean with parameters $(\lambda,a,b)$ implies stability w.r.t. the mean-deviation on $\mathcal{L}^1$ and the mean-upper-semideviation on $\mathcal{L}^1$, both with parameters $(\lambda,a(1+2\beta),b(1+2\beta))$. In the case of the mean-deviation, $\beta \in [0,\frac{1}{2}]$. In the case of the mean-upper-semideviation, $\beta \in [0,1]$.
\item Stability w.r.t. the mean-deviation on $\mathcal{L}^q$ implies stability w.r.t. the mean-upper-semideviation on $\mathcal{L}^q$ with the same parameters.
\item Stability w.r.t. $\varrho$ in region $S$ with parameters $(\lambda,a,b)$ implies the distributionally robust stability property w.r.t. the mean
\begin{equation*}
E(\psi(x_t) \xi) \leq a \lambda^t \psi(\mathbf{x}) + b, \quad \xi \in \mathcal{A}, \quad t \in \mathbb{N}, \quad \mathbf{x} \in S,
\end{equation*}
where $\mathcal{A}$ is the family of densities in the dual representation of $\varrho$.
\end{enumerate}
\end{theorem}
\begin{remark}[Interpretation of Theorem \ref{th1}]
The first item of Theorem \ref{th1} states that a CVaR stability property 
guarantees a probabilistic stability property. While connections between CVaR and probabilities are well-established, the first item of Theorem \ref{th1} is illuminating in light of the extensive history of probabilistic stability theory, e.g., see \cite{kushner1967stochastic, kannan2001handbook, meyn1993markov}.
%
%
The second and third items indicate that stability with respect to the mean ensures stability with respect to some common real-valued coherent risk functionals with transformed scale and offset parameters. The nature of the transformation depends on the specific family of densities in the dual representation. 
%
While the mean-deviation and the mean-upper-semideviation are similar functionals, the fourth item indicates that stability with respect to the mean-deviation is a stronger property. The last item states that stability with respect to a real-valued coherent risk functional on $\mathcal{L}^q$ guarantees a distributionally robust risk-neutral stability property, which we will see again in the final part of Theorem \ref{th2} (to be presented in Section \ref{lin}).  
\end{remark}

The proof of Theorem \ref{th1} has five parts, one for each item. Each part is based on the properties enjoyed by the risk functional at hand. 

\hspace{-7mm}\begin{proof}
Part 1: Let $t \in \mathbb{N}$ and $\mathbf{x} \in S$ be given. Since $\alpha \in (0,1)$ and $\psi(x_t) \in \mathcal{L}^1$, a minimizer of the right side of \eqref{cvardefdef} with $Z = \psi(x_t)$ is $\text{VaR}_{\alpha}(\psi(x_t))$ \cite[p. 258]{shapiro2009lectures}, and consequently,
$ \text{VaR}_{\alpha}(\psi(x_t)) \leq  \text{CVaR}_{\alpha}(\psi(x_t)) $.
Hence, $\text{CVaR}_{\alpha}(\psi(x_t)) \leq a \lambda^t \psi(\mathbf{x}) + b$ implies that
\begin{equation}
    \text{VaR}_{\alpha}(\psi(x_t)) \leq a \lambda^t \psi(\mathbf{x}) + b.
\end{equation}
Since $\alpha \in (0,1)$ and $z = a \lambda^t \psi(\mathbf{x}) + b \in \mathbb{R}$,
\begin{equation}
    \text{VaR}_{\alpha}(\psi(x_t)) \leq z \iff P(\{\psi(x_t) \leq z\}) \geq  1-\alpha
\end{equation}
by \cite[Lemma 21.1 (i)]{van2000asymptotic} 
and the definition of $\text{VaR}_{\alpha}$ \eqref{myvar}.

The following statement applies to both Parts 2 and 3, in which stability with respect to the mean is assumed. By Definition \ref{def1} and nonnegativity of $\psi$, it holds that $0 \leq E(\psi(x_t)) \leq a \lambda^t \psi(\mathbf{x}) + b$ for every $t \in \mathbb{N}$ and $\mathbf{x} \in S$. In particular, $\psi(x_t) \in \mathcal{L}^1$ for every $t \in \mathbb{N}$. 

Part 2: Let $t \in \mathbb{N}$ and $\mathbf{x} \in S$ be given. Since $\psi(x_t) \in \mathcal{L}^1$ and $\psi(x_t)$ is nonnegative, $\text{CVaR}_{\alpha}(\psi(x_t)) \leq \frac{1}{\alpha}E(\psi(x_t))$ \cite[Lemma 2]{mpctac2021}. Thus, $\text{CVaR}_{\alpha}(\psi(x_t)) \leq  \frac{1}{\alpha}(a \lambda^t \psi(\mathbf{x}) + b)$.

Part 3: For $\beta \in [0,\frac{1}{2}]$, the mean-deviation on $\mathcal{L}^1$ is real-valued and coherent. Denoting the associated family of densities by $\mathcal{A}$, every $\xi \in \mathcal{A}$ satisfies $0 \leq \xi \leq 1 + 2\beta$ a.e. (Remark \ref{remark1}). Let $t \in \mathbb{N}$, $\mathbf{x} \in S$, and $\xi \in \mathcal{A}$ be given. Since $\psi(x_t) \geq 0$ and $\xi \in [0, 1 + 2\beta]$ a.e., $0 \leq E(\psi(x_t) \xi) \leq E(\psi(x_t))(1 + 2\beta)$. Using $\psi(x_t) \in \mathcal{L}^1$ and taking the supremum over $\mathcal{A}$, we find that $\text{MD}_{1,\beta}(\psi(x_t)) \leq E(\psi(x_t))(1 + 2\beta)$. Then, the result follows from $0 \leq E(\psi(x_t)) \leq a \lambda^t \psi(\mathbf{x}) + b$. The derivation in the case of the mean-upper-semideviation 
is analogous.

Part 4: 
%
Let $\beta \in [0,\infty)$ be given. By assumption, $\psi(x_t) \in \mathcal{L}^q$ for every $t \in \mathbb{N}$, and there exist $\lambda \in [0,1)$, $a \in [0,\infty)$, and $b \in \mathbb{R}$ such that $$\text{MD}_{q,\beta}(\psi(x_t)) \leq a \lambda^t \psi(\mathbf{x}) + b$$ for every $t \in \mathbb{N}$ and $\mathbf{x} \in S$. Now, let $t \in \mathbb{N}$ and $\mathbf{x} \in S$ be given. Since $q \in [1,\infty)$, $\psi(x_t) \in \mathcal{L}^q$, and $P(\Omega) = 1$, $E(\psi(x_t))$ and $\|\psi(x_t) - E(\psi(x_t))\|_q$ are finite. Also, since $0 \leq \max\{g,0\} \leq |g|$ for any $g : (\Omega,\mathcal{F}) \rightarrow (\mathbb{R},\mathcal{B}_{\mathbb{R}})$ and $y \mapsto y^\gamma$ is nondecreasing on $[0,\infty)$ for any $\gamma \in (0,\infty)$,
\begin{equation*}
    \|\max\{\psi(x_t) - E(\psi(x_t)),0\}\|_q \leq \|\psi(x_t) - E(\psi(x_t))\|_q.
\end{equation*}
Hence, $\text{MUS}_{q,\beta}(\psi(x_t)) \leq \text{MD}_{q,\beta}(\psi(x_t)) \leq a \lambda^t \psi(\mathbf{x}) + b$.

Part 5: By assumption, $\psi(x_t) \in \mathcal{L}^q$ for every $t \in \mathbb{N}$, and there exist $\lambda \in [0,1)$, $a \in [0,\infty)$, and $b \in \mathbb{R}$ such that $\varrho(\psi(x_t)) \leq a \lambda^t \psi(\mathbf{x}) + b$ for every $t \in \mathbb{N}$ and $\mathbf{x} \in S$. Using the dual representation \eqref{dualrep} of $\varrho$ with the corresponding family $\mathcal{A}$ of densities and $\psi(x_t) \in \mathcal{L}^q$ for every $t \in \mathbb{N}$,
\begin{equation*}
   E(\psi(x_t) \xi) \leq \sup_{\xi \in \mathcal{A}} E(\psi(x_t) \xi) \overset{\eqref{dualrep}}{=} \varrho(\psi(x_t)) \leq a \lambda^t \psi(\mathbf{x}) + b
\end{equation*}
for every $\xi \in \mathcal{A}$, $t \in \mathbb{N}$, and $\mathbf{x} \in S$.
\end{proof}

The previous theorem provides connections between some different instances of Definition \ref{def2}. 
In the next section, we will focus on linear systems and derive sufficient conditions for stability with respect to any real-valued coherent risk functional on $\mathcal{L}^q$ (Theorem \ref{th2}). Then, we will provide a stability result in the context of the mean-conditional-variance functional on $\mathcal{L}^1$ (Theorem \ref{th3}). 
The techniques will involve a combination of the properties of the risk functionals of interest and Lyapunov stability theory for discrete-time linear systems.

\section{Risk-aware stability conditions\\for stochastic linear systems}\label{lin}
Now, we consider a stochastic linear time-invariant system, which is a special case of \eqref{nonlinsys}, of the form
\begin{equation}\label{my88}
    x_{t+1} = A x_t + w_t, \quad t = 0,1, \dots,
\end{equation}
where $A \in \mathbb{R}^{n \times n}$ is deterministic, $(x_0,x_1,\dots)$ is an $\mathbb{R}^n$-valued stochastic process, and $(w_0, w_1,\dots)$ is an $\mathbb{R}^n$-valued independent stochastic process. The processes are defined on $(\Omega,\mathcal{F},P)$. 
The initial state is fixed at an arbitrary vector $\mathbf{x} \in \mathbb{R}^n$. 
%
%
For every $t \in \mathbb{N}_0$, we define the random object $h_t \coloneqq (x_0,x_1,\dots,x_t)$ and the $\sigma$-algebra $\mathcal{F}_t \coloneqq \sigma(h_t)$ induced by $h_t$.
%
%
Thus, $x_t : (\Omega,\mathcal{F}_t) \rightarrow (\mathbb{R}^n,\mathcal{B}_{\mathbb{R}^n})$. 
%
As well as $(w_0, w_1,\dots)$ being independent, we assume that $w_t$ and $h_t$ are independent and $E(w_t w_t^\top) \in \mathbb{R}^{n \times n}$ for every $t \in \mathbb{N}_0$. 
%
%
%
The above conditions are standard and ensure that $E(|x_t|^2)$ is finite for every $t \in \mathbb{N}_0$. Observe that $(w_0, w_1,\dots)$ need not be zero-mean, and the disturbance process need not be identically distributed. When we consider the linear system \eqref{my88}, we implicitly assume the conditions stated in this paragraph. 
%
%
%

The first result of the section will develop sufficient conditions for stability of the linear system \eqref{my88} with respect to any real-valued coherent risk functional on $\mathcal{L}^q$. Then, we will examine how the conditions are related to those that guarantee risk-neutral stability of \eqref{my88}.

\begin{theorem}[Coherent stability]\label{th2}
Consider the linear system \eqref{my88} and the quadratic state energy function $\psi(x) \coloneqq x^\top R x$, where $R \in \mathcal{S}_n^+$ is given. Let $\varrho$ be a real-valued coherent risk functional on $\mathcal{L}^q$ with $q \in [1,\infty)$, where $\mathcal{A}$ is the family of densities in the associated dual representation. We make the following assumptions:
\begin{enumerate}
    \item $|w_t|^2 \in \mathcal{L}^q$ for every $t \in \mathbb{N}_0$, $\underset{t \in \mathbb{N}_0}{\sup} \big \|  |w_t|^2  \big \|_q$ is finite, and
    \item there exists an $H \in \mathcal{S}_n^+$ such that 
    $H_R - A^\top H_R A \in \mathcal{S}_n^+$. 
    %
  %
\end{enumerate}
(Recall that $H_R \coloneqq (R^{\frac{1}{2}})^\top H R^{\frac{1}{2}}$.) Then, the system \eqref{my88} is stable with respect to $\varrho$ in $\mathbb{R}^n$ (Definition \ref{def2}). Define $\eta \coloneqq \frac{\lambda_{\text{min}}(H_R - A^\top H_R A)}{\lambda_{\text{max}}(H_R)}$, and it follows that $\eta \in (0,1]$. In the case of $\eta < 1$, for any fixed $\kappa\in(0,1)$, one can choose 
\begin{equation}\label{myparamth2}
  \textstyle  \lambda  = 1 - \kappa\eta , \quad \quad a  =  \frac{\lambda_{\text{max}}(H)}{\lambda_{\text{min}}(H)} , \quad \quad b = c b',
    %
\end{equation}
where $\lambda \in (0,1)$, $c \coloneqq    \frac{\lambda}{\lambda_{\text{min}}(H) (1-\lambda) (\lambda - (1-\eta))}$, and
\begin{equation}\label{mybprime}
  b' \coloneqq \sup_{t \in \mathbb{N}_0} \varrho(w_t^\top H_R w_t).
\end{equation}
In the case of $\eta = 1$, one can choose $\lambda = a = 0$ and $b= \bar{c} b'$, where $b'$ is defined by \eqref{mybprime} and $\bar{c} \coloneqq 1/\lambda_{\text{min}}(H)$.
%
%
Moreover, a distributionally robust risk-neutral stability property holds:
 $  E(\psi(x_t) \xi) \leq a  \lambda^t  \psi(\mathbf{x}) + b $ 
for every density $\xi \in \mathcal{A}$, time $t \in \mathbb{N}$, and initial condition $\mathbf{x} \in \mathbb{R}^n$.
\end{theorem}




\begin{remark}[Discussion of Theorem \ref{th2}]
Theorem \ref{th2} specifies sufficient conditions that guarantee
\begin{equation}\label{2222}
    \varrho(x_t^\top R x_t) \leq a \lambda^t \mathbf{x}^\top R \mathbf{x} + \tilde{c} \sup_{t \in \mathbb{N}_0} \varrho(w_t^\top H_R w_t)
\end{equation}
for every $t \in \mathbb{N}$ and $\mathbf{x} \in \mathbb{R}^n$, where $a \in [0,\infty)$, $\lambda \in [0,1)$, $\tilde{c} \in (0,\infty)$, and $H_R \in \mathcal{S}_n^+$ are constant. ($\tilde{c} = c$ or $\tilde{c} = \bar{c}$, where $c$ and $\bar{c}$ are provided by Theorem \ref{th2}.) The statement \eqref{2222} is a risk-aware noise-to-state stability property, where $\varrho$ evaluates a noise ``energy'' term $w_t^\top H_R w_t$ and the state energy $x_t^\top R x_t$. Moreover, since $\lambda \in [0,1)$, we derive the following risk-aware stability bound:
\begin{equation}\label{242424}
    \limsup_{t \rightarrow \infty} \varrho(x_t^\top R x_t) \leq \tilde{c} \sup_{t \in \mathbb{N}_0} \varrho(w_t^\top H_R w_t).
\end{equation}

Now, let us discuss the assumptions of Theorem \ref{th2}. In the case of $q = 1$, $\sup_{t \in \mathbb{N}_0} \big \|  |w_t|^2  \big \|_1 = \sup_{t \in \mathbb{N}_0} E(|w_t|^2)$ is finite, for example, if there is a matrix $\Sigma \in \mathcal{S}_n$ such that $\Sigma - E(w_t w_t^\top) \in \mathcal{S}_n$ for every $t \in \mathbb{N}_0$.
If $E(w_t w_t^\top)$ is time-invariant, then such a $\Sigma$ exists immediately. (Recall that $E(w_t w_t^\top) \in \mathbb{R}^{n \times n}$ in the model \eqref{my88}.) More generally, the suitability of the first assumption is problem-dependent, and the assumption holds in particular if $w_t$ is uniformly bounded almost everywhere.
Since $R \in \mathcal{S}_n^+$, the second assumption 
is equivalent to $A$ being Schur stable; i.e., every eigenvalue of $A$ has magnitude strictly less than one (Lemma \ref{lemma2appendix}, Appendix). 


We have developed our proof of Theorem \ref{th2} from first principles using several techniques, including properties of $\mathcal{S}_n$ from \cite{horn2012matrix}, some basic manipulations from \cite[Lemma 1]{morozan1983stabilization}, 
and properties of real-valued coherent risk functionals from \cite{shapiro2009lectures}. 
\end{remark}
\hspace{-5mm}
\begin{proof}
The property $\eta \in (0,1]$ holds by Lemma \ref{etalemma} (Appendix). 
We shall prove the theorem in the case of $\eta < 1$ and leave the case of $\eta = 1$ to the reader, as the techniques are similar. First, we will verify some properties of the parameters in \eqref{myparamth2}. The properties $\eta \in (0,1)$ and $\kappa \in (0,1)$ 
imply that $\lambda  \in (0,1)$.
The property $a \in (0,\infty)$ holds because $H \in \mathcal{S}_n^+$. The property $c \in (0,\infty)$ is true because $\lambda \in (0,1)$, $\lambda > 1-\eta$, and $H \in \mathcal{S}_n^+$. For every $\xi \in \mathcal{A}$ and $t \in \mathbb{N}_0$, 
\begin{equation}\label{18}
    0 \leq E(w_t^\top H_R w_t \xi) \leq \lambda_{\max}(H_R) E(|w_t|^2 \xi) 
\end{equation}
because $H_R \in \mathcal{S}_n^+$, $w_t$ is $\mathbb{R}^n$-valued, and $\xi \geq 0$ a.e. 
Also,
\begin{equation}\label{2424}
    E(|w_t|^2 \xi) \leq \| \, |w_t|^2 \, \|_q \; \|\xi\|_{q*}, \quad \xi \in \mathcal{A}, \quad t \in \mathbb{N}_0,
\end{equation}
by applying \cite[H\"{o}lder's Inequality, Th. 6.8 (a)]{folland1999real}. 
Since $\varrho$ is a real-valued coherent risk functional on $\mathcal{L}^q$, $\mathcal{A}$ is a bounded subset of $\mathcal{L}^{q*}$ \cite[Th. 6.6]{shapiro2009lectures}. The property $b \in [0,\infty)$ follows from \eqref{18}--\eqref{2424}, boundedness of $\mathcal{A}$ in $\| \cdot \|_{q*}$, the assumptions of the theorem, the dual representation of $\varrho$ \eqref{dualrep}, and $c \in (0,\infty)$. 
%
%

Second, we will verify properties of some expectations.
For any density $\xi \in \mathcal{A}$, initial condition $\mathbf{x} \in \mathbb{R}^n$, and function $g : (\mathbb{R}^n,\mathcal{B}_{\mathbb{R}^n}) \rightarrow (\mathbb{R},\mathcal{B}_{\mathbb{R}})$,
\begin{equation}\label{19}
    E(g(x_0)\xi) = g(\mathbf{x})
\end{equation}
because $E(\xi) = 1$ and the distribution of $x_0$ is the Dirac measure concentrated at $\mathbf{x}$. 
%
%
%
Let $t \in \mathbb{N}_0$, $\xi \in \mathcal{A}$, and $M \in \mathcal{S}_n$ be given. Then, $x_t^\top M x_t \in \mathcal{L}^q$ and $x_t^\top M x_t \xi \in \mathcal{L}^1$, and these functions are a.e.-nonnegative (Lemma \ref{quadformlemma}, Appendix). 
In particular, since $R \in \mathcal{S}_n^+$ and $H_R \in \mathcal{S}_n^+$, $\psi(x_t) = x_t^\top R x_t \in \mathcal{L}^q$ and $E(x_t^\top H_R x_t \xi) \in [0,\infty)$.
 
Having verified the above properties, we are ready to show that $\varrho(\psi(x_t)) \leq a \lambda^t \psi(\mathbf{x}) + b$ for every time $t \in \mathbb{N}$ and initial condition $\mathbf{x} \in \mathbb{R}^n$. Define $v : \mathbb{R}^n \rightarrow \mathbb{R}$ by $v(z) \coloneqq z^\top H_R z$, which will serve as a Lyapunov function. 
%
%
Recall the definition of $b'$ from \eqref{mybprime}.
We claim it suffices to prove that, for every $\xi \in \mathcal{A}$ and $t \in \mathbb{N}_0$, 
\begin{equation}\label{20}
    \textstyle E(v(x_{t+1})\xi) \leq \lambda E(v(x_{t})\xi) + \frac{\lambda}{\lambda - (1-\eta)}b'.
\end{equation}
%
%
Indeed, the statement \eqref{20} and $E(v(x_{t})\xi) \in [0,\infty)$ for every $t \in \mathbb{N}_0$ imply 
\begin{equation}\label{21}
    0 \leq E(v(x_{t})\xi) \leq \lambda^t E(v(x_0) \xi) + \textstyle \frac{\lambda b'}{(1-\lambda)(\lambda - (1-\eta))} 
\end{equation}
for every $t \in \mathbb{N}$, where we use $\lambda \in (0,1)$, $\lambda > 1 - \eta$, the geometric series formula, and $b' \in [0,\infty)$ (Lemma \ref{geom}, Appendix). Since $H \in \mathcal{S}_n^+$, $R \in \mathcal{S}_n^+$, and every $\xi \in \mathcal{A}$ is nonnegative a.e., we have
\begin{equation}\label{statement22}
    \hspace{-6bp}\lambda_{\text{min}}(H) E(\psi(x_t)\xi) \leq E(v(x_t)\xi) \leq \lambda_{\text{max}}(H) E(\psi(x_t)\xi) \hspace{-3bp}
\end{equation}
%
%
for every $\xi \in \mathcal{A}$ and $t \in \mathbb{N}_0$ (Lemma \ref{quadformlemma}, Appendix). We divide by $\lambda_{\text{min}}(H) \in (0,\infty)$ to find
\begin{equation}\label{23}
   \textstyle 0 \leq E(\psi(x_t)\xi) \leq \frac{1}{\lambda_{\text{min}}(H)}E(v(x_t)\xi).
\end{equation}
Using \eqref{statement22} with $t = 0$ and \eqref{19} with $g = \psi$ lead to
\begin{equation}\label{24}
    \hspace{-6bp}0 \leq E(v(x_0)\xi) \leq \lambda_{\text{max}}(H) E(\psi(x_0)\xi) = \lambda_{\text{max}}(H) \psi(\mathbf{x}) \hspace{-3bp}
\end{equation}
for every density $\xi \in \mathcal{A}$ and initial condition $\mathbf{x} \in \mathbb{R}^n$. 
The statements \eqref{21} and \eqref{24} imply 
\begin{equation}\label{25}
    0 \leq E(v(x_{t})\xi) \leq \lambda^t \lambda_{\text{max}}(H) \psi(\mathbf{x}) + \textstyle \frac{\lambda b'}{(1-\lambda)(\lambda - (1-\eta))} 
\end{equation}
for every $\xi \in \mathcal{A}$, $t \in \mathbb{N}$, and $\mathbf{x} \in \mathbb{R}^n$. We divide \eqref{25} by $\lambda_{\text{min}}(H) \in (0,\infty)$ and substitute the expressions for $a$ and $b$ from 
\eqref{myparamth2} to find
\begin{equation}\label{27}
   \textstyle 0 \leq \frac{1}{\lambda_{\text{min}}(H)} E(v(x_{t})\xi) \leq a \lambda^t \psi(\mathbf{x}) + b
\end{equation}
for every $\xi \in \mathcal{A}$, $t \in \mathbb{N}$, and $\mathbf{x} \in \mathbb{R}^n$. The statements \eqref{23} and \eqref{27} imply 
\begin{equation}\label{31}
   0 \leq E(\psi(x_t)\xi) \leq a \lambda^t \psi(\mathbf{x}) + b
\end{equation}
for every $\xi \in \mathcal{A}$, $t \in \mathbb{N}$, and $\mathbf{x} \in \mathbb{R}^n$.
Then, taking the supremum over $\mathcal{A}$ and using the dual representation \eqref{dualrep} of $\varrho$ lead to
\begin{equation}
    \varrho(\psi(x_t)) \leq a \lambda^t \psi(\mathbf{x}) + b, \quad t \in \mathbb{N}, \quad \mathbf{x} \in \mathbb{R}^n.
\end{equation}
Hence, it suffices to show \eqref{20}, which we prove in the Appendix (see Lemma \ref{keyu} and Lemma \ref{show20}).
\end{proof}

Theorem \ref{th2} provides sufficient interpretable conditions for global stability of the linear system \eqref{my88} with respect to any real-valued coherent risk functional $\varrho$ on $\mathcal{L}^q$ (Definition \ref{def2}). We have presented some common examples of $\varrho$ in Section \ref{nonlin}, which assess different distributional characteristics of the state energy. 
For instance, $\varrho(\psi(x_t)) = \text{CVaR}_{\alpha}(\psi(x_t))$ is the average energy in the $\alpha$-fraction of the largest energies, 
if $\psi(x_t)$ is a continuous random variable.
For another example, $\varrho(\psi(x_t)) = \text{MUS}_{q,\beta}(\psi(x_t))$ is a weighted sum of the average energy $\mu_t \coloneqq E(\psi(x_t))$ and the order-$q$ upper-semideviation $\| \max\{\psi(x_t) - \mu_t,0\} \|_q$ of the energy. 
Theorem \ref{th2} facilitates the analysis of various characteristics of the distribution of the state energy
%
%
and 
offers a distributionally robust stability guarantee (see also Part 5 of Theorem \ref{th1}). 
%

It is quite interesting that the assumptions of Theorem \ref{th2} in the case of $q = 1$ ensure 
risk-neutral stability of the system \eqref{my88}, i.e., 
the existence of $\lambda_{\text{N}} \in [0,1)$, $a_{\text{N}} \in [0,\infty)$, and $b_{\text{N}} \in \mathbb{R}$ such that 
\begin{equation}
    E(\psi(x_t)) \leq a_{\text{N}} \lambda_{\text{N}}^t \psi(\mathbf{x}) + b_{\text{N}}, \quad t \in \mathbb{N}, \quad \mathbf{x} \in \mathbb{R}^n.
\end{equation}
While the proof of this risk-neutral result is nontrivial, it is a simpler version of the proof of Theorem \ref{th2} (consider $\xi = 1$ a.e. and $q = 1$).
The values of $\lambda_{\text{N}}$ and $a_{\text{N}}$ can be chosen to be the values of $\lambda$ and $a$, respectively, provided by Theorem \ref{th2}. However, the offset term $b_{\text{N}}$ in the risk-neutral result and the offset term $b$ in the risk-aware result (Theorem \ref{th2}) are distinct.
In the case of $\eta < 1$, one can choose
\begin{equation}\label{mymy22}
    b_{\text{N}} = c \sup_{t \in \mathbb{N}_0} E(w_t^\top H_R w_t),
\end{equation}
while in the case of $\eta = 1$, one can choose
   $ b_{\text{N}} = \bar{c} \sup_{t \in \mathbb{N}_0} E(w_t^\top H_R w_t) $,
where $c$ and $\bar{c}$ are specified by Theorem \ref{th2}. 

Let us further discuss the different offset terms. The offset term $b_{\text{N}}$ in \eqref{mymy22} is proportional to the supremum of the mean of $w_t^\top H_R w_t$. In contrast, the offset term $b$ in Theorem \ref{th2} is proportional to the supremum of the risk of $w_t^\top H_R w_t$, where the meaning of risk is specific to the functional $\varrho$ of interest. As well as permitting a risk-aware perception of $w_t^\top H_R w_t$ that is specific to the primal interpretation of $\varrho$, 
Theorem \ref{th2}
permits distributional ambiguity in $w_t^\top H_R w_t$ due to the dual representation of $\varrho$. 
This attribute is useful in settings with \textcolor{black}{distributional modeling uncertainty, which we will illustrate by using the mean-upper-semideviation as an example (Section \ref{secV}, Illustration \ref{illus1})}. 
Theorem \ref{th3} to follow will use a mean-conditional-variance functional $\rho_\nu$ (Example \ref{ex4}) and additional assumptions about the distribution of $w_t$, which are inspired by \cite{tsiamis2020risk}. 
Theorem \ref{th3} will not provide a distributional robustness property, which is not surprising because $\rho_\nu$ is noncoherent in general. However, Theorem \ref{th3} will provide a versatile risk-aware noise-to-state stability property that depends on second-, third-, and fourth-order centered noise statistics, which is relevant for controller design (to be explored in Illustration \ref{illus3}, Section \ref{secV}). 
%
%
Prior to presenting Theorem \ref{th3}, we will present some useful notation. 
%
%
\begin{remark}[Notation for Theorem \ref{th3}]\label{remark4}
As in Theorem \ref{th2}, Theorem \ref{th3} will also involve the linear system \eqref{my88} and a quadratic state energy function $\psi(x) \coloneqq x^\top R x$, where $R \in \mathcal{S}_n^+$ is given. 
%
%
%
Since $\psi(x_t) \in \mathcal{L}^1$ for any $t \in \mathbb{N}$, the mean-conditional-variance of $\psi(x_t)$ is well-defined (Example \ref{ex4}). For every $t \in \mathbb{N}$, $\rho_\nu(\psi(x_t))$ denotes the mean-conditional-variance of $\psi(x_t)$, where $\mathcal{F}_{t-1} = \sigma(x_0,x_1,\dots,x_{t-1})$ is the sub $\sigma$-algebra of $\mathcal{F}$ of interest. For every $t \in \mathbb{N}$, we define $\bar{w}_{t-1} \coloneqq E(w_{t-1})$, $d_t \coloneqq w_{t-1} - \bar{w}_{t-1}$, and $\Sigma_t \coloneqq E(d_t d_t^\top)$. 
\end{remark}

\begin{theorem}[Mean-cond.-variance stability]\label{th3} Consider
the stochastic linear system \eqref{my88}, and let $R\in{\cal S}_{n}^{+}$ and 
$\nu \in [0,\infty)$ 
be given. Consider the quadratic state energy function $\psi(x)\coloneqq x^{\top}Rx$.
We make the following assumptions:
\begin{enumerate}
\item 
$E(|w_{t}|^{4})$ is finite for every $t\in\mathbb{N}_{0}$, and there exists a matrix $\Sigma_{u}\in{\cal S}_{n}$ such that $\Sigma_{u}-\Sigma_{t} \in \mathcal{S}_n$ for every $t\in\mathbb{N}$.
\item Define $R_{\nu}\coloneqq R+4\nu R\Sigma_{u}R$. There exists a matrix $H^{\nu}\in{\cal S}_{n}^{+}$ such that $H_{R_{\nu}}^{\nu}-A^{\top}H_{R_{\nu}}^{\nu}A\in{\cal S}_{n}^{+}$.
\item 
The statistics $\bar{{\bf w}}_{t}$, $\gamma_{t}$, and $\delta_{t}$ defined by
\begin{equation}\label{mystatnew}\begin{aligned}
\bar{{\bf w}}_{t} & \coloneqq \textstyle (I_{n}-A^{t+1})^{-1}\sum_{i=0}^{t}A^{i}\bar{w}_{t-i}, && t \in \mathbb{N}_0,\\
\gamma_{t} & \coloneqq E(d_{t}d_{t}^{\top}Rd_{t}) , && t\in\mathbb{N},\\
\delta_{t} & \coloneqq E((d_{t}^{\top}Rd_{t}-\text{tr}(\Sigma_{t}R))^{2}), && t\in\mathbb{N},
\end{aligned}\end{equation}
satisfy $\sup_{t\in\mathbb{N}_{0}}|\bar{{\bf w}}_{t}|^{2} < \infty$, $\sup_{t\in\mathbb{N}}|\gamma_{t}|^{2}<\infty$, and
$\sup_{t\in\mathbb{N}}\delta_{t}<\infty$, respectively.
 %
%
\end{enumerate}
Then, 
there exist $\{\lambda_{\nu},\lambda_{0}\} \subset [0,1)$,  $\{ a_{\nu}, a_0 \} \subset [0,\infty)$,
and $b_{\nu}\in\mathbb{R}$ such that
\begin{align}
 & \rho_{\nu}(\psi(x_{t}))\nonumber \\
 & \le\underbrace{a_{\nu}\lambda_{\nu}^{t}\tilde{\psi}_{\nu}({\bf x})+4\nu|\gamma_{t}|\sqrt{a_{0}\lambda_{0}^{t}  \lambda_{\text{max}}(R)\tilde{\psi}_{0}({\bf x})  }}_{\text{Exponentially decreasing terms}} \; + \; b_{\nu} \label{my4141}
\end{align}
\textcolor{black}{for every time $t \in \mathbb{N}$ and initial condition $\mathbf{x} \in \mathbb{R}^n$,}
where we define $\tilde{\psi}_{\nu}({\bf x})\hspace{-1bp} \coloneqq \hspace{-1bp}4\psi_{\nu}({\bf x})+4\lambda_{\text{max}}(R_{\nu})\sup_{t\in\mathbb{N}_{0}}|\bar{{\bf w}}_{t}|^{2}$ and $\psi_{\nu}({\bf x})\hspace{-1bp}\hspace{-1bp} \coloneqq \hspace{-1bp}{\bf x}^{\top}R_{\nu}{\bf x}$.
In particular, one can choose
\begin{equation}
\lambda_{\nu}=1 -{ \frac{\lambda_{\text{min}}(H_{R_{\nu}}^{\nu}-A^{\top}H_{R_{\nu}}^{\nu}A)}{\lambda_{\text{max}}(H_{R_{\nu}}^{\nu})}} 
,\;\;\;\;a_{\nu}={ \frac{\lambda_{\text{max}}(H^{\nu})}{\lambda_{\text{min}}(H^{\nu})}}  ,
\end{equation}
and
\begin{align}\label{mybnu}
    b_{\nu} & = \sup_{t \in \mathbb{N}} \big\{ \rho_{\nu}(\psi(d_{t})) + b_{\nu,t} \big\},
\end{align}
where, for every $t \in \mathbb{N}$, 
\begin{align}
b_{\nu,t} & \coloneqq \underbrace{c_{\nu}\Big ( \sup_{t\in\mathbb{N}}E(d_{t}^{\top}H_{R_{\nu}}^{\nu}d_{t})\Big ) -E(d_{t}^{\top}R_{\nu,t}d_{t})}_{\text{Risk-neutral zero-mean noise bias}}\nonumber \\
 & \quad  +\underbrace{2\bar{{\bf w}}_{t-1}^{\top}R_{\nu}\bar{{\bf w}}_{t-1}+4\nu\gamma_{t}^{\top}R\bar{{\bf w}}_{t-1}}_{\text{Noise mean and skewness terms}}, \label{mybnut}
\end{align}
$R_{\nu,t}\coloneqq R+4\nu R\Sigma_{t}R$, and $c_{\nu}\coloneqq\frac{2}{\lambda_{\text{min}}(H^{\nu})(1-\lambda_{\nu})}$.
%
%
\end{theorem}\medskip{}

\begin{remark}[Discussion of Theorem \ref{th3}] 
Theorem \ref{th3} establishes a global stability property 
with respect to $\rho_\nu$, closely resembling Definition \ref{def2}, where a sum of exponentially decreasing terms appears on the right side of \eqref{my4141}. 
Similar to Theorem \ref{th2}, Theorem \ref{th3} also uncovers a risk-aware noise-to-state stability property. 
%
By denoting the sum of exponentially decreasing terms in \eqref{my4141} by $g_{\nu,\mathbf{x},t}$ and substituting the expression for $b_{\nu}$ from \eqref{mybnu}, we can express \eqref{my4141} as follows:
\begin{equation}\label{45}
    \rho_\nu(\psi(x_{t})) \leq g_{\nu,\mathbf{x},t} + \sup_{t \in \mathbb{N}} \big\{ \rho_{\nu}(\psi(d_{t})) + b_{\nu,t} \big\} 
\end{equation}
for every time $t \in \mathbb{N}$ and initial condition $\mathbf{x} \in \mathbb{R}^n$. The risk functional $\rho_\nu$ evaluates the state energy on the left side of \eqref{45}, and $\rho_\nu$ evaluates the energy of the centered noise $d_t$ on the right side of \eqref{45}.
%
Under the assumptions of Theorem \ref{th3}, $g_{\nu,\mathbf{x},t}$ decays to zero as $t \rightarrow \infty$, and therefore, Theorem \ref{th3} provides the following risk-aware stability bound:
\begin{equation}\label{my4747}
   \limsup_{t \rightarrow \infty}  \rho_\nu(\psi(x_{t})) \leq \sup_{t \in \mathbb{N}} \big\{ \rho_{\nu}(\psi(d_{t})) + b_{\nu,t} \big\}.
\end{equation}

Now, let us discuss the assumptions of Theorem \ref{th3}. 
For any random variable $Y$ on $(\Omega,\mathcal{F},P)$, $E(Y^4)$ is finite in many cases,  including when $Y$ has a Gaussian distribution, a distribution with bounded support, a log-normal distribution (which is heavy-tailed), or a mixture of these distributions. 
Using $R_\nu \in \mathcal{S}_n^+$, the second assumption of Theorem \ref{th3} is equivalent to $A$ being Schur stable (Lemma \ref{lemma2appendix}, Appendix). Hence, one is not an eigenvalue of $A$, which ensures that $I_n - A^{t+1}$ is invertible for every $t \in \mathbb{N}_0$. In the case of $\bar{w}_{t} = \bar{w}$ for every $t \in \mathbb{N}_0$, it follows that $\bar{{\bf w}}_{t}=(I_{n}-A)^{-1}\bar{w}$, and so $\sup_{t\in\mathbb{N}_{0}}|\bar{{\bf w}}_{t}|^{2}$ is finite (more details to be provided in the proof). We will prove Theorem \ref{th3} next.
\end{remark}

\hspace{-6mm}\begin{proof}Let $\nu \in [0,\infty)$ be given, and let $\mathbf{x} \in \mathbb{R}^n$ be an arbitrary initial condition of the system \eqref{my88}. Unrolling the
recursion 
provides $x_{t+1}=A^{t+1}x_0+\sum_{i=0}^{t}A^{i}w_{t-i}$ for every $t\in\mathbb{N}_{0}$,
and thus, 
\begin{align}
x_{t+1} 
 & =A^{t+1}x_0+\sum_{i=0}^{t}A^{i}\bar{w}_{t-i}+\sum_{i=0}^{t}A^{i}(w_{t-i}-\bar{w}_{t-i})\nonumber \\
 & =A^{t+1}x_0+(I_{n}-A^{t+1})\bar{{\bf w}}_{t}+\sum_{i=0}^{t}A^{i}d_{t-i+1},
\end{align}
where the first line of \eqref{mystatnew} specifies the definition of $\bar{{\bf w}}_{t}$. 
That is, for every $t \in \mathbb{N}_0$, we have the ``centered'' solution
\begin{equation}
x_{t+1}-\bar{{\bf w}}_{t}=A^{t+1}(x_0-\bar{{\bf w}}_{t})+\sum_{i=0}^{t}A^{i}d_{t-i+1}.\label{eq:CENTERED}
\end{equation}
It is convenient to define the ``centered'' state
\begin{equation}
\hat{x}_{t+1}\coloneqq x_{t+1}-\bar{{\bf w}}_{t}\iff x_{t+1}=\hat{x}_{t+1}+\bar{{\bf w}}_{t} \label{eq:HAT}
\end{equation}
for every $t \in \mathbb{N}_0$.
Additionally, the mean of $x_{t+1}$ is given by 
\begin{equation}
E(x_{t+1})=A^{t+1}{\bf x}+\sum_{i=0}^{t}A^{i}\bar{w}_{t-i}=A^{t+1}({\bf x}-\bar{{\bf w}}_{t})+\bar{{\bf w}}_{t} \label{eq:MEAN}
\end{equation}
for every $t \in \mathbb{N}_0$. As a special case, in the setting of mean-stationary noise, i.e., $\bar{w}_{t} = \bar{w}$ for every $t \in \mathbb{N}_0$, we obtain
\begin{align}
E(x_{t+1}) & =A^{t+1}{\bf x}+\sum_{i=0}^{t}A^{i}\bar{w}\nonumber \\
 & =A^{t+1}{\bf x}+(I_{n}-A^{t+1})(I_{n}-A)^{-1}\bar{w},
\end{align}
and so $\bar{{\bf w}}_{t}=(I_{n}-A)^{-1}\bar{w}$.
($A$ being Schur stable 
implies that $\sum_{i=0}^{t}A^{i}=(I_{n}-A^{t+1})(I_{n}-A)^{-1}$ \cite[Ex. 5.6.P26]{horn2012matrix}.) 
This case may be analyzed as well and provide 
improved constants, but we do not pursue it further for the sake of generality.

Now, let us study the stability of the system with respect to $\rho_{\nu}$. For convenience, we define $R_{\nu}\coloneqq R+4\nu R\Sigma_{u}R$ and $\mathbb{C}_{t}$ for every $t \in \mathbb{N}$ by
\begin{align}
\mathbb{C}_{t} & \coloneqq \nu\delta_{t}-4\nu\text{tr}((\Sigma_{t}R)^{2})\nonumber \\
 & \hphantom{:} =\nu{\textbf{var}}(d_{t}^{\top}Rd_{t})-4\nu E(d_{t}^{\top}R\Sigma_{t}Rd_{t}), \label{my555555}
\end{align}
where $\textbf{var}$ denotes variance. For every $t \in \mathbb{N}$, we have (see Lemma \ref{bigquadform} in the Appendix for the first line) 
\begin{align}
\rho_{\nu}(\psi(x_{t})) & =E(x_{t}^{\top}(R+4\nu R\Sigma_{t}R)x_{t}+4\nu x_{t}^{\top}R\gamma_{t})+\mathbb{C}_{t}\nonumber \\
 & =E(x_{t}^{\top}(R+4\nu R\Sigma_{t}R)x_{t})+4\nu\gamma_{t}^{\top}RE(x_{t})+\mathbb{C}_{t}\nonumber \\
 & \le E(x_{t}^{\top}(R+4\nu R\Sigma_{u}R)x_{t})+4\nu\gamma_{t}^{\top}RE(x_{t})+\mathbb{C}_{t}\nonumber \\
 & =E(x_{t}^{\top}R_{\nu}x_{t})+4\nu\gamma_{t}^{\top}RE(x_{t})+\mathbb{C}_{t}, \label{my5656}
\end{align}
where we have used the assumption that $\Sigma_u \in \mathcal{S}_n$ satisfies $\Sigma_u - \Sigma_t \in \mathcal{S}_n$ for every $t \in \mathbb{N}$. 

The risk $\rho_{\nu}(\psi(x_{t}))$ has two distinct dynamical terms
whose stability requires analysis, 
i.e., the \textit{quadratic
term} $E(x_{t}^{\top}R_{\nu}x_{t})$ and the \textit{cross-term} $\gamma_{t}^{\top}RE(x_{t})$.
For the quadratic term, we use (\ref{eq:HAT}) to write, for every
$t\in\mathbb{N}$, 
\begin{equation}\label{57}
E(x_{t}^{\top}R_{\nu}x_{t})\le2E(\hat{x}_{t}^{\top}R_{\nu}\hat{x}_{t})+2\bar{{\bf w}}_{t-1}^{\top}R_{\nu}\bar{{\bf w}}_{t-1}.
\end{equation}
For the expectation on the right side of \eqref{57}, we apply (\ref{eq:CENTERED}) to derive (note that the noise process $d_{t}$ is zero-mean)
\begin{align}
 & E(\hat{x}_{t}^{\top}R_{\nu}\hat{x}_{t})\nonumber \\
 & =E\bigg((x_0-\bar{{\bf w}}_{t-1})^{\top}[A^{t}]^{\top}R_{\nu}A^{t}(x_0-\bar{{\bf w}}_{t-1})\nonumber \\
 & \quad\quad+\bigg[\sum_{i=0}^{t-1}A^{i}d_{t-i}\bigg]^{\top}R_{\nu}\bigg[\sum_{i=0}^{t-1}A^{i}d_{t-i}\bigg]\bigg)\nonumber \\
 & \le E\bigg ( 2x_0^{\top}[A^{t}]^{\top}R_{\nu}A^{t}x_0+2\bar{{\bf w}}_{t-1}^{\top}[A^{t}]^{\top}R_{\nu}A^{t}\bar{{\bf w}}_{t-1}\nonumber \\
 & \quad\quad+\bigg[\sum_{i=0}^{t-1}A^{i}d_{t-i}\bigg]^{\top}R_{\nu}\bigg[\sum_{i=0}^{t-1}A^{i}d_{t-i}\bigg]\bigg ) \nonumber \\
 & =E(\tilde{x}_{t}^{\top}R_{\nu}\tilde{x}_{t})+2\bar{{\bf w}}_{t-1}^{\top}[A^{t}]^{\top}R_{\nu}A^{t}\bar{{\bf w}}_{t-1}, \label{my58}
\end{align}
for every $t \in \mathbb{N}$, where the process $(\tilde{x}_{1},\tilde{x}_{2},\dots)$ is defined by
\begin{equation}
\tilde{x}_{t+1}=A^{t+1}\sqrt{2}x_0+\sum_{i=0}^{t}A^{i}d_{t-i+1}, \quad t \in \mathbb{N}_0.
\end{equation}
Equivalently, it holds that 
\begin{equation}
\tilde{x}_{t+1}=A\tilde{x}_{t}+d_{t+1},\quad t \in \mathbb{N}_0, 
\end{equation}
where we define $\tilde{x}_{0} \coloneqq \sqrt{2}x_0$. Now, consider a Schur stable deterministic system defined by $z_{t+1} = Az_{t}$ for every $t \in \mathbb{N}_0$ with an arbitrary initial state $z_{0}=\boldsymbol{z} \in \mathbb{R}^n$. It holds that\footnote{For any $t \in \mathbb{N}$, it holds that 
\begin{align}
\bar{{\bf w}}_{t-1}^{\top}[A^{t}]^{\top}R_{\nu}A^{t}\bar{{\bf w}}_{t-1}\nonumber 
 & =\big|R_{\nu}^{1/2}A^{t}\bar{{\bf w}}_{t-1}\big|^{2}\nonumber \\
 & \le\Big(\big|R_{\nu}^{1/2}A^{t}\big|_{2} \; \big|\bar{{\bf w}}_{t-1}\big|\Big)^{2}\nonumber \\
 & =\big|\bar{{\bf w}}_{t-1}\big|^{2}\bigg(\sup_{|\boldsymbol{z}|=1}\big|R_{\nu}^{1/2}A^{t}\boldsymbol{z}\big|\bigg)^{2}\nonumber \\
 & =\big|\bar{{\bf w}}_{t-1}\big|^{2}\sup_{|\boldsymbol{z}|=1}\big|R_{\nu}^{1/2}A^{t}\boldsymbol{z}\big|^{2}\nonumber \\
 & =|\bar{{\bf w}}_{t-1}|^{2}\sup_{|\boldsymbol{z}|=1}\boldsymbol{z}^{\top}[A^{t}]^{\top}R_{\nu}A^{t}\boldsymbol{z}\nonumber \\
 & =|\bar{{\bf w}}_{t-1}|^{2}\sup_{|\boldsymbol{z}|=1}z_{t}^{\top}R_{\nu}z_{t}.
\end{align}
}
\begin{equation}
    0  \le\bar{{\bf w}}_{t-1}^{\top}[A^{t}]^{\top}R_{\nu}A^{t}\bar{{\bf w}}_{t-1} \le |\bar{{\bf w}}_{t-1}|^{2}\sup_{|\boldsymbol{z}|=1}z_{t}^{\top}R_{\nu}z_{t} 
\end{equation}
for every $t \in \mathbb{N}$. Consequently, we have shown that 
\begin{equation}
E(\hat{x}_{t}^{\top}R_{\nu}\hat{x}_{t})\le E(\tilde{x}_{t}^{\top}R_{\nu}\tilde{x}_{t})+2|\bar{{\bf w}}_{t-1}|^{2}\sup_{|\boldsymbol{z}|=1}z_{t}^{\top}R_{\nu}z_{t}
\end{equation}
for every $t \in \mathbb{N}$. Next, let us study the cross term $\gamma_{t}^{\top}RE(x_{t})$ in \eqref{my5656}. For any $t \in \mathbb{N}$, we use (\ref{eq:MEAN}) to find
\begin{align}
 & \,\gamma_{t}^{\top}RE(x_{t})\\
 & =\hspace{-1bp}\gamma_{t}^{\top}RA^{t}({\bf x}-\bar{{\bf w}}_{t-1})+\gamma_{t}^{\top}R\bar{{\bf w}}_{t-1}\nonumber \\
 & \le\hspace{-1bp}|R^{1/2}\gamma_{t}||R^{1/2}A^{t}({\bf x}-\bar{{\bf w}}_{t-1})|+\gamma_{t}^{\top}R\bar{{\bf w}}_{t-1}\nonumber \\
 & =\hspace{-1bp}|R^{1/2}\gamma_{t}|\sqrt{({\bf x}-\bar{{\bf w}}_{t-1})^{\top}[A^{t}]^{\top}RA^{t}({\bf x}-\bar{{\bf w}}_{t-1})}\nonumber \\
 & \quad\quad+\gamma_{t}^{\top}R\bar{{\bf w}}_{t-1}\nonumber \\
 & \le\hspace{-1bp}|R^{1/2}\gamma_{t}|\sqrt{2{\bf x}^{\top}[A^{t}]^{\top}RA^{t}{\bf x}+2\bar{{\bf w}}_{t-1}^{\top}[A^{t}]^{\top}RA^{t}\bar{{\bf w}}_{t-1}}\nonumber \\
 & \quad\quad+\gamma_{t}^{\top}R\bar{{\bf w}}_{t-1}\nonumber \\
 & \le\hspace{-1bp}|R^{1/2}\gamma_{t}|\sqrt{[x_{t}^{d}]^{\top}Rx_{t}^{d}\hspace{-1bp}+\hspace{-1bp}2|\bar{{\bf w}}_{t-1}|^{2}\sup_{|\boldsymbol{z}|=1}z_{t}^{\top}Rz_{t}}+\hspace{-1bp}\gamma_{t}^{\top}R\bar{{\bf w}}_{t-1},\nonumber 
\end{align}
where $x_{t+1}^{d}  = Ax_{t}^{d}$ for every $t \in \mathbb{N}_0$ with initialization $x_{0}^{d}=\sqrt{2}{\bf x}$
is another Schur stable deterministic system. 
Overall,
for every $t\in\mathbb{N}$, we have shown that
\begin{align}
\rho_{\nu}(\psi(x_{t})) & \le2E(\tilde{x}_{t}^{\top}R_{\nu}\tilde{x}_{t})+4|\bar{{\bf w}}_{t-1}|^{2}\sup_{|\boldsymbol{z}|=1}z_{t}^{\top}R_{\nu}z_{t}\nonumber \\
 & \;\;+4\nu|R^{1/2}\gamma_{t}|\sqrt{[x_{t}^{d}]^{\top}Rx_{t}^{d}+2|\bar{{\bf w}}_{t-1}|^{2}\sup_{|\boldsymbol{z}|=1}z_{t}^{\top}Rz_{t}}\nonumber \\
 & \;\;+2\bar{{\bf w}}_{t-1}^{\top}R_{\nu}\bar{{\bf w}}_{t-1}+4\nu\gamma_{t}^{\top}R\bar{{\bf w}}_{t-1}+\mathbb{C}_{t} .\label{eq:Bound_1}
\end{align}
To complete the proof, we will analyze the terms in the right side of (\ref{eq:Bound_1}).
%
%
For convenience, we define
\begin{equation}\label{66}
    s_\nu \coloneqq \sup_{t\in\mathbb{N}}E(d_{t}^{\top}H_{R_{\nu}}^{\nu}d_{t}), \quad \eta_\nu \coloneqq \dfrac{\lambda_{\text{min}}(H_{R_{\nu}}^{\nu}-A^{\top}H_{R_{\nu}}^{\nu}A)}{\lambda_{\text{max}}(H_{R_{\nu}}^{\nu})},
\end{equation}
and $\lambda_\nu \coloneqq 1 - \eta_\nu$, and thus, $s_\nu \in [0,\infty)$ and $\lambda_\nu  \in [0,1)$ using the assumptions of Theorem \ref{th3} (also see Lemma \ref{etalemma}, Appendix). Moreover, we define $\psi_{\nu}(y) \coloneqq y^{\top}R_{\nu}y$ for every $y \in \mathbb{R}^n$ and $a_\nu \coloneqq \lambda_{\text{max}}(H^{\nu})/\lambda_{\text{min}}(H^{\nu})$; note that $a_\nu \in (0,\infty)$ by the second assumption. Using $R \in \mathcal{S}_n^+$ and $R_\nu \in \mathcal{S}_n^+$, the second assumption implies the existence of a matrix $H^0 \in \mathcal{S}_n^+$ such that $H^0_{R} - A^\top H_{R}^0 A \in \mathcal{S}_n^+$ (see Lemma \ref{lemma2appendix}, Appendix). 
We define $R_0 \coloneqq R$, $a_0 \coloneqq \lambda_{\text{max}}(H^0)/\lambda_{\text{min}}(H^0)$, $\eta_0$ by the analogous expression in \eqref{66} with $\nu = 0$, $\lambda_0 \coloneqq 1 - \eta_0$, and $\psi_{0}(y) \coloneqq y^{\top}R_{0}y$ for every $y \in \mathbb{R}^n$. 
By applying the second assumption in particular, the following statements about the terms in (\ref{eq:Bound_1}) hold:
\begin{itemize}
\item \textit{ The term \mbox{$E(\tilde{x}_{t}^{\top}R_{\nu}\tilde{x}_{t})$}}: 
\begin{align}
  \hspace{-1bp}\hspace{-1bp}\hspace{-1bp}\hspace{-1bp}\hspace{-1bp}\hspace{-1bp}E(\tilde{x}_{t}^{\top}R_{\nu}\tilde{x}_{t}) 
  \le 2 a_\nu \lambda_{\nu}^{t}\psi_{\nu}({\bf x})
  + \dfrac{s_\nu}{\lambda_{\text{min}}(H^{\nu})(1-\lambda_{\nu})}, \; \; t \in \mathbb{N}. \label{my70}
\end{align}
\item \textit{The term \mbox{$z_{t}^{\top}R_{\nu}z_{t}$} for every \mbox{$\nu\ge0$}}: 
\begin{align}
z_{t}^{\top}R_{\nu}z_{t} & \le a_\nu\lambda_{\nu}^{t}\psi_{\nu}({\bf z}), \quad t \in \mathbb{N},  \quad \mathbf{z} \in \mathbb{R}^n,
\end{align}
and therefore, for every $t \in \mathbb{N}$,
\begin{align}
\sup_{|\boldsymbol{z}|=1}z_{t}^{\top}R_{\nu}z_{t} & \le a_\nu\lambda_{\nu}^{t}\sup_{|\boldsymbol{z}|=1}\psi_{\nu}({\bf z}) 
  =a_\nu\lambda_{\text{max}}(R_{\nu})\lambda_{\nu}^{t} .
\end{align}
\item \textit{The term \mbox{$[x_{t}^{d}]^{\top}Rx_{t}^{d}$}}:
 \begin{align}
[x_{t}^{d}]^{\top}Rx_{t}^{d}\le  2 a_0 \lambda_{0}^{t}\psi_{0}({\bf x}), \quad t \in \mathbb{N}.
\end{align} 
\end{itemize}
By defining
\begin{align}
c_{\nu} & \coloneqq  \dfrac{2}{\lambda_{\text{min}}(H^{\nu})(1-\lambda_{\nu})}, \\
\tilde{\psi}_{\nu}({\bf x}) & \coloneqq4\psi_{\nu}({\bf x})+4\lambda_{\text{max}}(R_{\nu})\sup_{t\in\mathbb{N}}|\bar{{\bf w}}_{t-1}|^{2}, \label{my7171} 
\end{align}
and $\tilde{\psi}_0({\bf x})$ by \eqref{my7171} with $\nu = 0$, 
we consolidate our previous 
results as follows:
\begin{align}\label{my7373}
 %
 2E(\tilde{x}_{t}^{\top}R_{\nu}\tilde{x}_{t})+4|\bar{{\bf w}}_{t-1}|^{2}\sup_{|\boldsymbol{z}|=1}z_{t}^{\top}R_{\nu}z_{t} 
  \le a_{\nu}\lambda_{\nu}^{t}\tilde{\psi}_{\nu}({\bf x})+c_{\nu} s_\nu  
\end{align}
and
\begin{align}\label{my747474}
 [x_{t}^{d}]^{\top}Rx_{t}^{d}+2|\bar{{\bf w}}_{t-1}|^{2}\sup_{|\boldsymbol{z}|=1}z_{t}^{\top}Rz_{t} 
  \le a_{0}\lambda_{0}^{t}\tilde{\psi}_{0}({\bf x})
\end{align}
for every $t \in \mathbb{N}$.
The statements \eqref{eq:Bound_1}, \eqref{my7373}, and \eqref{my747474} provide the following (almost final) bound for any $t \in \mathbb{N}$:
\begin{align}
\rho_{\nu}(\psi(x_{t})) & \le\underbrace{a_{\nu}\lambda_{\nu}^{t}\tilde{\psi}_{\nu}({\bf x})+4\nu|R^{1/2}\gamma_{t}|\sqrt{a_{0}\lambda_{0}^{t}\tilde{\psi}_{0}({\bf x})}}_{\text{Exponentially decreasing terms}} \label{almost}\\
 & \quad\;\underbrace{+ \; c_{\nu} s_\nu +2\bar{{\bf w}}_{t-1}^{\top}R_{\nu}\bar{{\bf w}}_{t-1}+4\nu\gamma_{t}^{\top}R\bar{{\bf w}}_{t-1}+\mathbb{C}_{t}}_{\text{Bias terms}}.\nonumber 
\end{align}
Next, we use \eqref{my555555}, the definition $R_{\nu,t}\coloneqq R+4\nu R\Sigma_{t}R$, and the independence of $d_t$ and $h_{t-1}$ for any $t \in \mathbb{N}$ to derive 
\begin{align}
\mathbb{C}_{t} 
 & =\nu{ \textbf{var}}(d_{t}^{\top}Rd_{t})+E(d_{t}^{\top}Rd_{t})-E(d_{t}^{\top}Rd_{t})\nonumber \\
 & \quad\quad-4\nu E(d_{t}^{\top}R\Sigma_{t}Rd_{t})\nonumber \\
 & =\underbrace{E(d_{t}^{\top}Rd_{t})+\nu{\textbf{var}}(d_{t}^{\top}Rd_{t})}_{\text{\text{Noise mean-variance}}}-E(d_{t}^{\top}R_{\nu,t}d_{t})\nonumber \\
 & =\rho_{\nu}(\psi(d_{t}))-E(d_{t}^{\top}R_{\nu,t} d_{t}) . \label{almostalmost}
\end{align}
Finally, we use $|R^{1/2}\gamma_{t}| \le\sqrt{\lambda_{\text{max}}(R)}|\gamma_{t}|$, \eqref{almost}, and \eqref{almostalmost} to derive
\begin{align}
 & \hspace{-1bp}\rho_{\nu}(\psi(x_{t}))\nonumber \\
 & \le\underbrace{a_{\nu}\lambda_{\nu}^{t}\tilde{\psi}_{\nu}({\bf x})+4\nu|\gamma_{t}|\sqrt{a_{0}\lambda_{0}^{t}  \lambda_{\text{max}}(R)\tilde{\psi}_{0}({\bf x})  }}_{\text{Exponentially decreasing terms}}\nonumber \\
 & \quad\quad+\underbrace{c_{\nu}\Big ( \sup_{t\in\mathbb{N}}E(d_{t}^{\top}H_{R_{\nu}}^{\nu}d_{t})\Big ) -E(d_{t}^{\top}R_{\nu,t}d_{t})}_{\text{Risk-neutral zero-mean noise bias}}\nonumber \\
 & \quad\quad\quad+\underbrace{\rho_{\nu}(\psi(d_{t}))}_{\text{Noise risk}}+\underbrace{2\bar{{\bf w}}_{t-1}^{\top}R_{\nu}\bar{{\bf w}}_{t-1}+4\nu\gamma_{t}^{\top}R\bar{{\bf w}}_{t-1}}_{\text{Noise mean and skewness terms}}
\end{align}
for every $t\in\mathbb{N}$, completing the proof. \end{proof}

\begin{figure*}
  \includegraphics[scale = 0.446]{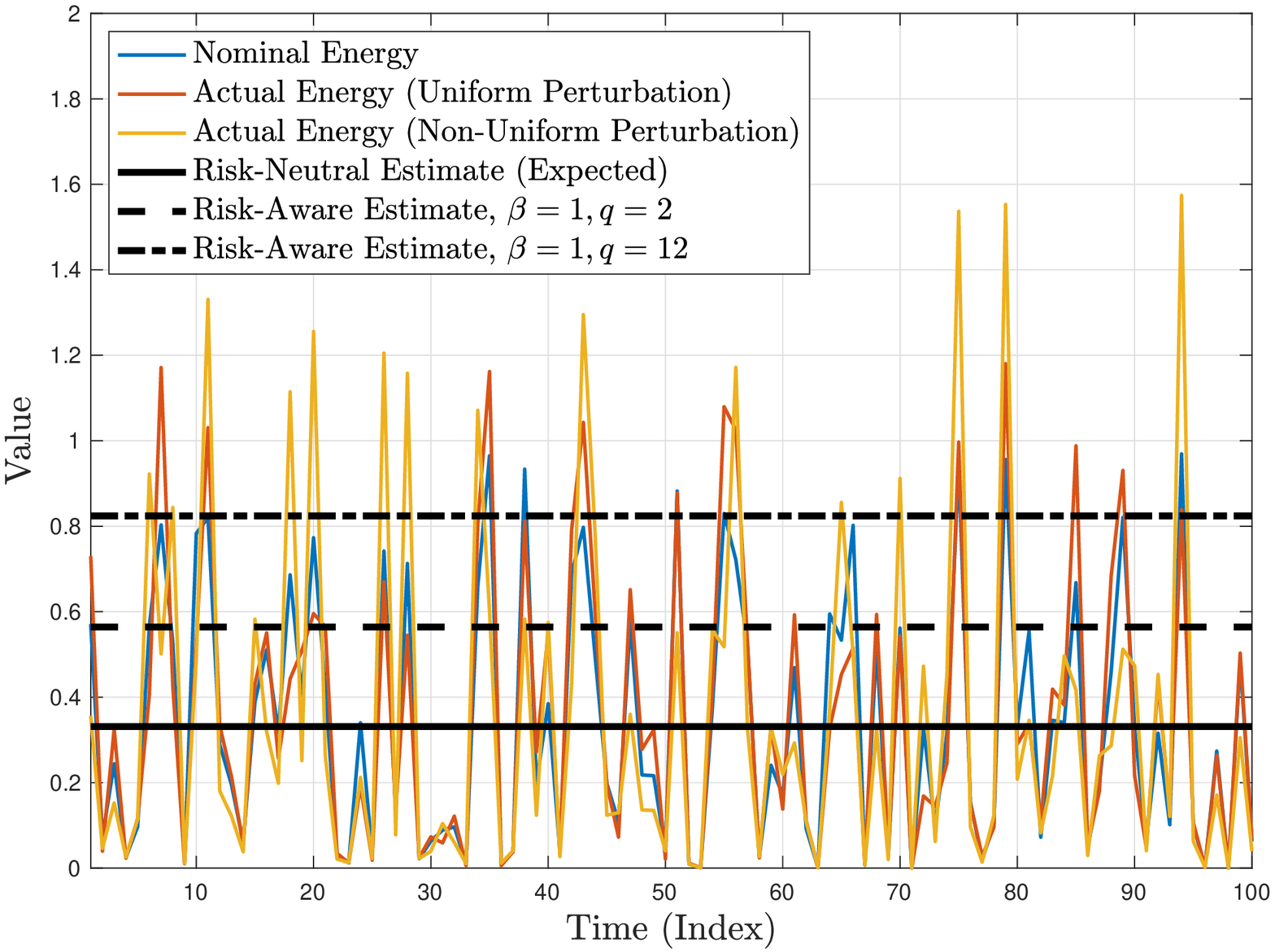}
  \hspace{6bp}
  \includegraphics[scale = 0.446]{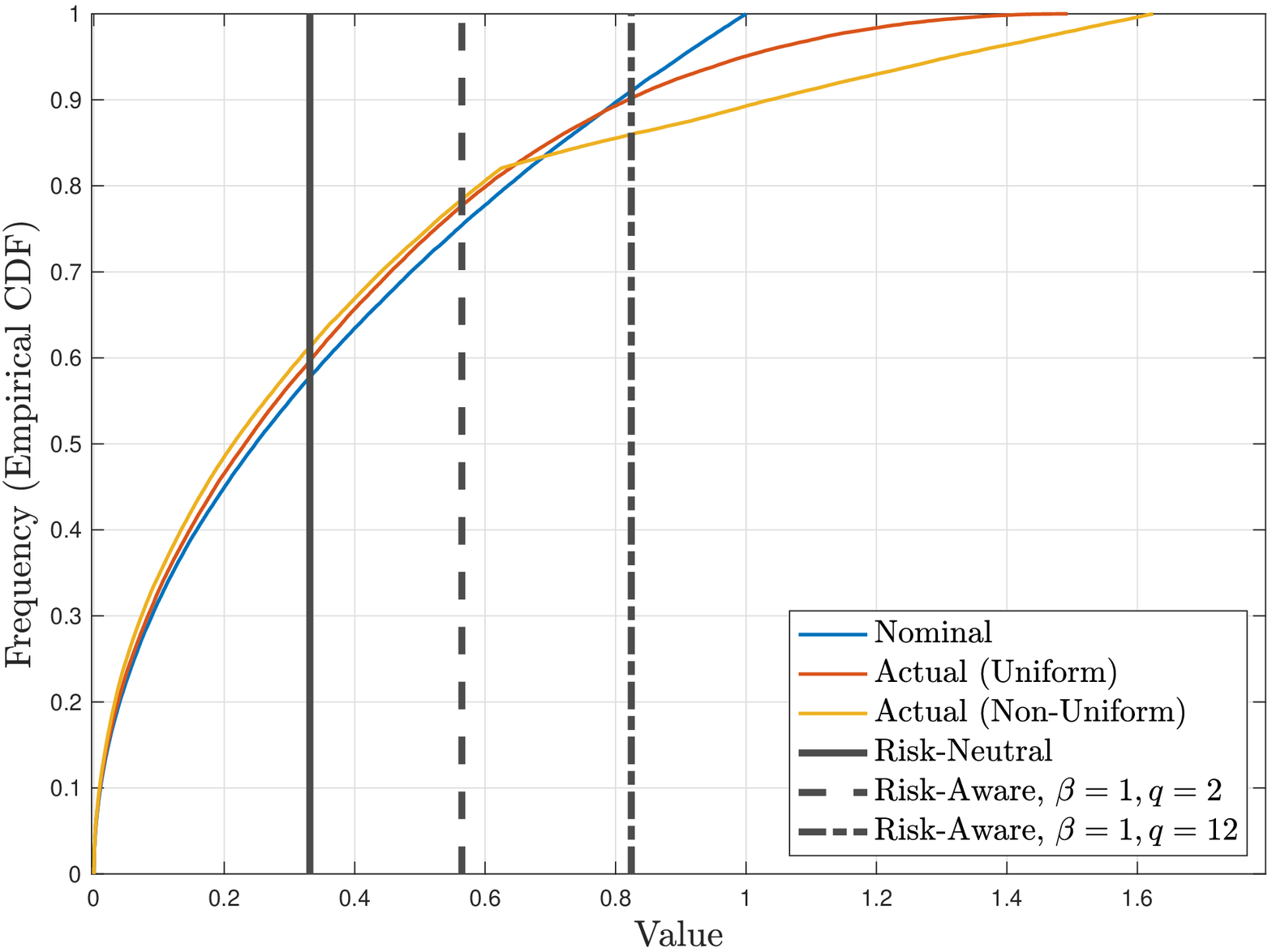}
  \caption{\label{Realities} Time trajectories (left) and respective empirical CDFs (right) of the noise energy under the nominal model and the two alternative realities described in Illustration \ref{illus1}. The plots also show the obtained risk-neutral and risk-aware noise energy estimates (straight horizontal \& vertical lines).}
  \vspace{-10pt}
\end{figure*}
%

Similarities and differences between Theorem \ref{th2} and Theorem \ref{th3} deserve some discussion. Both theorems show how the risk of the state energy and a risk-aware perception of a disturbance process can be decoupled, although the theorems quantify risk in distinct ways; compare \eqref{2222} and \eqref{45}. Theorem \ref{th2} applies to any real-valued coherent risk functional on $\mathcal{L}^q$, while Theorem \ref{th3} applies to a mean-conditional-variance functional on $\mathcal{L}^1$.
The offset parameters from the two theorems 
each reflect a particular perception of noise, which is induced by the chosen measure of risk; indeed, compare \eqref{mybprime} and \eqref{mybnu}. 
Theorem \ref{th3} provides a rate parameter $\lambda_\nu$, which depends on $\nu$, where $\nu$ is specific to the risk functional $\rho_\nu$. However, the rate parameter from Theorem \ref{th2} is equivalent to the one in the risk-neutral setting. 
Later on, we will see that Theorem \ref{th3} provides insights about the behavior of a simple risk-aware myopic controller, which mitigates extreme peaks of the state energy in simulation (Section \ref{secV}, Illustration \ref{illus3}). In contrast, how to develop a simple risk-aware controller in the setting of Theorem \ref{th2} is an open research question.

Ideally, we would like to optimize the parameters provided by Theorem \ref{th2} or Theorem \ref{th3} by choosing a suitable state-feedback matrix $K$. For example, one can consider the problem of minimizing the offset parameter $b = c  \sup_{t \in \mathbb{N}_0} \varrho(w_t^\top H_R w_t)$ \eqref{myparamth2} subject to the assumptions of Theorem \ref{th2}, where $A = \check{A} - \check{B}K$, $\check{A}$ and $\check{B}$ are given matrices, and $K$ is to be chosen. This is a difficult risk-aware optimization problem to solve exactly, where the matrix variables $H$ and $K$ are coupled through $A^\top H_R A = (\check{A} - \check{B}K)^\top H_R (\check{A} - \check{B}K)$. 
Moreover, the rate parameter $\lambda$ \eqref{myparamth2} should be minimized simultaneously, adding to the difficulty.  
Hence, we reserve investigations of such optimization problems for future work. The next section will provide examples to illuminate aspects of our theoretical developments. 


\section{Illustrative Examples}\label{secV}

\begin{illusexample}[Risk of noise energy]\label{illus1}
Theorem \ref{th2} indicates that the risk of the state energy $x_t^\top R x_t$ of the system under study is of the order of the corresponding risk of the noise energy $w_t^\top H_R w_t$ under some assumptions. (The term \emph{noise energy} refers to any quadratic form $w_t^\top M w_t$ with $M \in \mathcal{S}_n^+$.) 
We consider for simplicity
a one-dimensional, Schur stable linear system driven by independent and identically
distributed noise $w_{t}$ (i.e., a stable order-$1$ autoregressive
process). This illustration will examine the risk of $w_t^2$, which arises in the right side of $b'$ \eqref{mybprime} in a special case, compared to realizations of $w_t^2$ to demonstrate
the need for and the usefulness of evaluating the noise energy 
from a risk-aware perspective.

Let us select the mean-upper-semideviation of order-$q$ \eqref{my7} 
for this illustration. 
Note that
whenever $\beta\in[0,1]$ we obtain a coherent risk functional for every value of $q\ge1$, with its
dual representation \eqref{dualrep} 
being true with the uncertainty set ${\cal A}$ being given by \eqref{7}. 

\begin{figure*}
  \includegraphics[scale = 0.446]{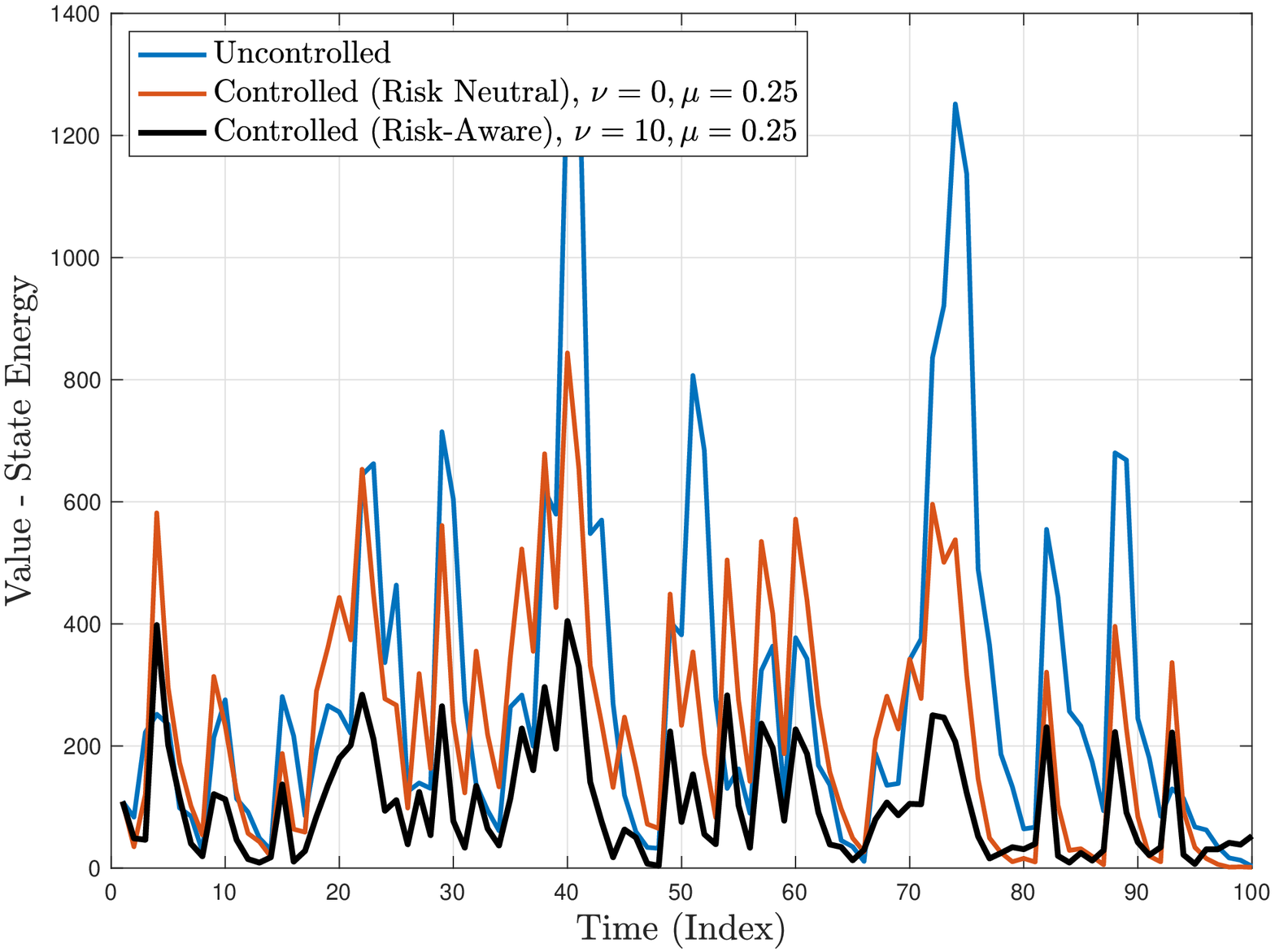}
  \hspace{6bp}
  \includegraphics[scale = 0.446]{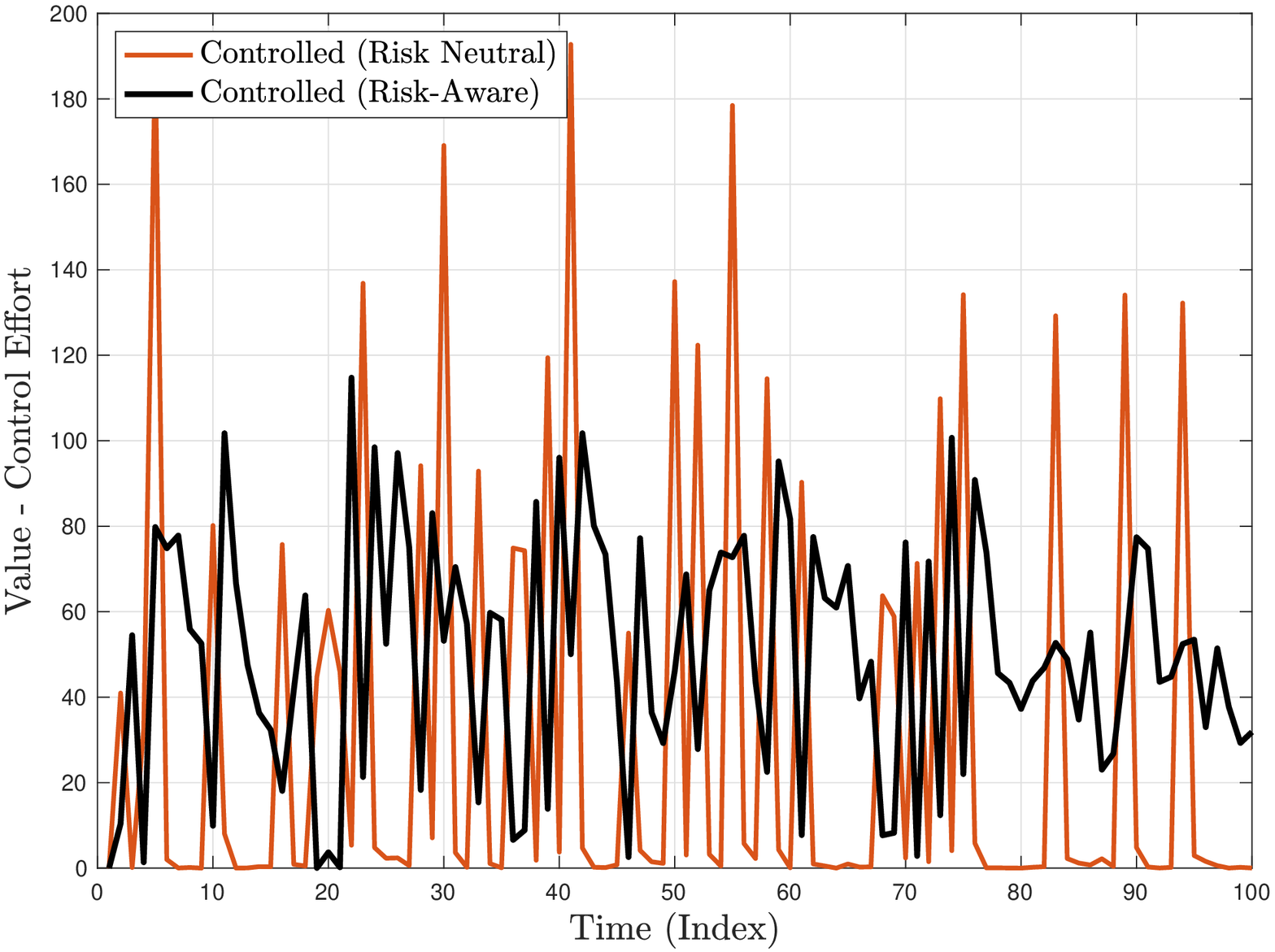}
  \\
  \includegraphics[scale = 0.446]{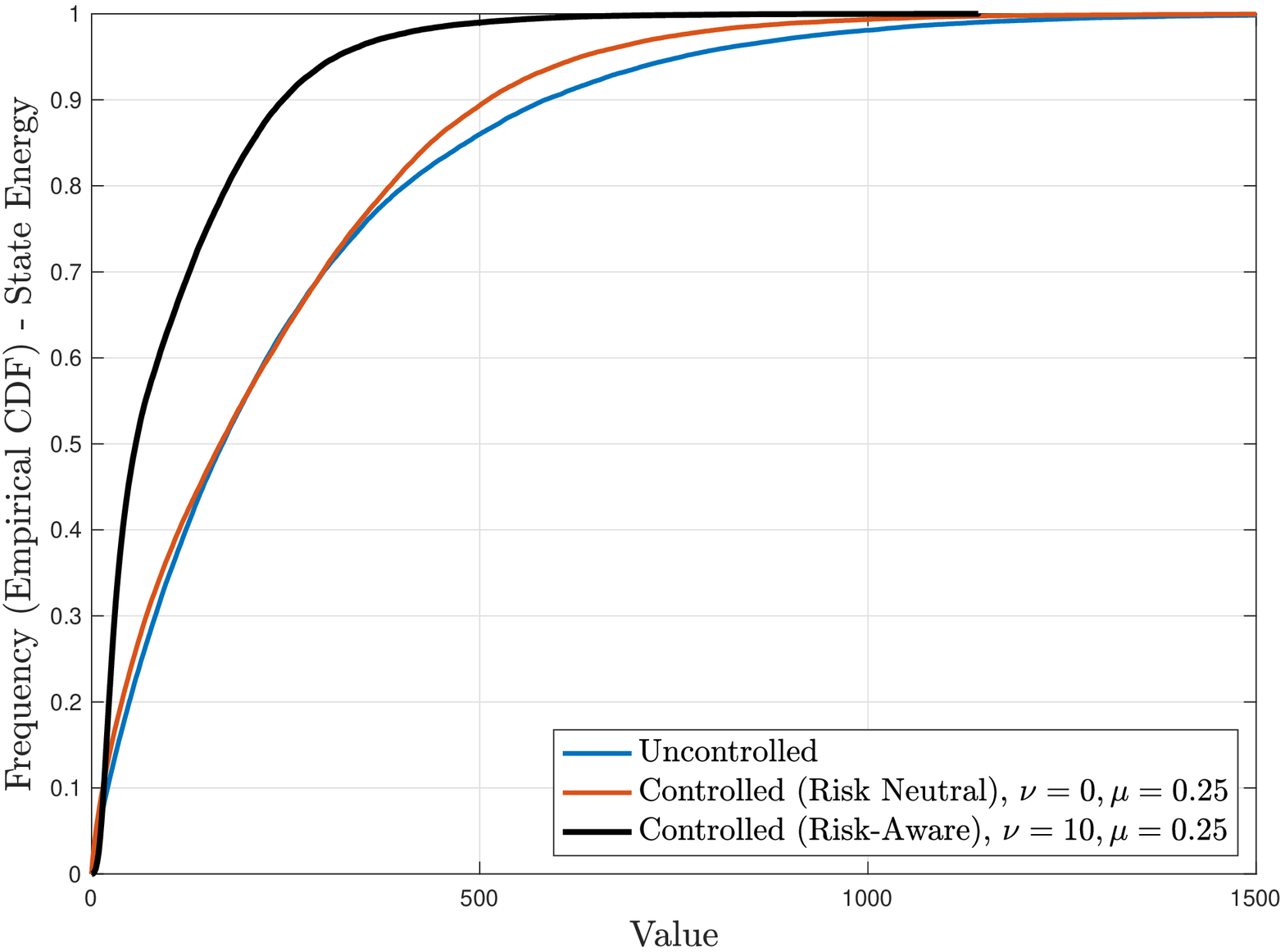}
  \hspace{6bp}
  \includegraphics[scale = 0.446]{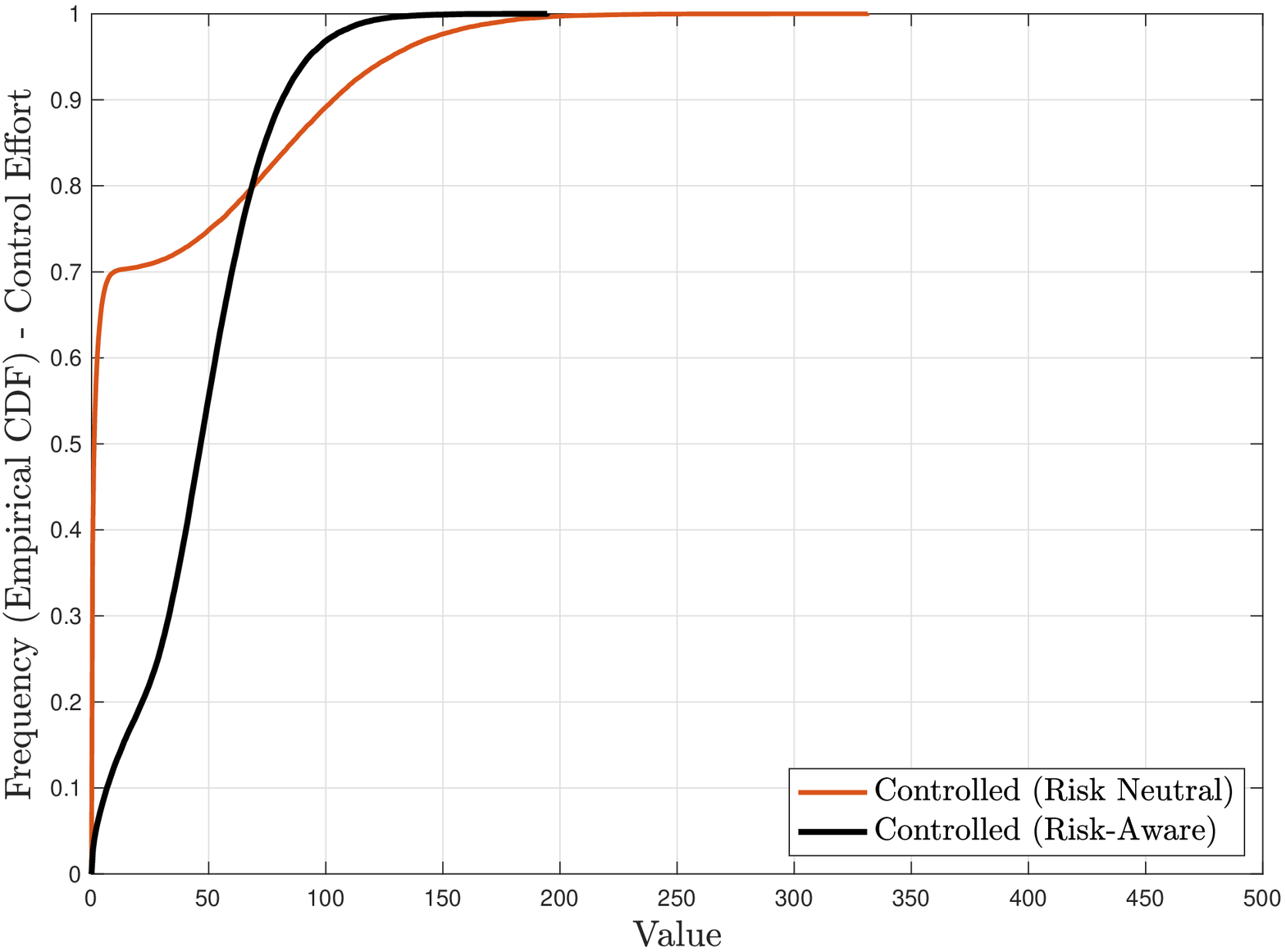}
  \caption{\label{RealitiesController} The time trajectories (top) and empirical CDFs (bottom) of
the state energy $|x_t|^2$ and the respective control effort $|u_t|^2$ (where applicable)
corresponding to no controller ($u_t$ is zero), the risk-neutral controller
for $\mu=0.25$, and the risk-aware controller for $\mu=0.25$ and
$\nu=10$ for the stochastic linear system described in Illustration \ref{illus3}. Despite the simplicity of the risk-aware controller, this controller can mitigate extreme peaks in the state energy without requiring extreme peaks in the control effort.}
  \vspace{-10pt}
\end{figure*}

While we may assume a generic \textit{nominal} noise model, e.g., $w_{t}\sim{\cal U}[-1,1]$, 
at the same time we may be uncertain about it.
In such a case, 
we would
like to obtain a \text{robust} estimate of the \text{average} energy
of the noise 
with respect to potential distributional modeling
uncertainty. 
%
%
This is precisely achieved by using a (coherent) risk functional 
to evaluate $w_{t}$ \text{under its nominal model} (which arises from characterizing system stability through Theorem
\ref{th2}). 
To illustrate this, let us consider two alternative
realities for $w_{t}^{2}$ defined by
\begin{align}
w_{t,i}^{2} & \coloneqq w_{t}^{2}\xi_{t,i} , \quad i\in\{1,2\},\quad t\in\mathbb{N}_0 ,
\end{align}
where $\xi_{t,i}$ for $i\in\{1,2\}$ 
are multiplicative random perturbation processes
of the form 
\begin{equation}
\xi_{t,i}=1+\eta_{t,i}-E(\eta_{t,i}),\quad i\in\{1,2\},
\end{equation}
where, for every $t\in\mathbb{N}_0$,
\begin{align}
\eta_{t,1} & \sim{\cal U}[0,1]\quad\text{and}\\
\eta_{2,t} & = z_t\mathcal{I}_{[0,1]}(z_t)+\mathcal{I}_{>1}(z_t),
\end{align}
%
%
with $z_t=-w_t+1/4$. In other words, each $\xi_{t,i}$ for $i\in\{1,2\}$
belongs to the uncertainty set \eqref{7} associated with the order-$q$ mean-upper-semideviation
risk functional \eqref{my7} 
for every $q\ge1$. Observe
that $\xi_{t,1}$ is a uniform perturbation and independent of $w_t$, whereas $\xi_{t,2}$ is nonuniform and highly dependent on $w_t$. Still, both $\xi_{t,1}$ and $\xi_{t,2}$ are \text{each} independent over different values of $t$ 
(and thus the same happens for $w^2_{t,i},i\in\{1,2\}$, which is helpful in numerical simulation).
%
%
It follows that, for $\beta\in[0,1]$ and $q\ge1$,
\begin{equation}
E(w_{t,i}^{2})\le\sup_{\xi\in{\cal A}}E(w_{t}^{2}\xi)=\mathrm{MUS}_{q,\beta}(w_{t}^{2}),\quad i\in\{1,2\},
\end{equation}
illustrating that the mean-upper-semideviation risk functional (which is just one example) can provide worst-case risk-neutral estimates of the noise energy over a particular but rich class of possible realities, including the ones described above.

Figure \ref{Realities} illuminates the above facts by depicting the time trajectories and corresponding empirical CDFs
of the noise energy 
in all three cases. Figure \ref{Realities} also shows
the risk-neutral estimate ($\beta=0$) and two risk-aware estimates ($\beta=1$, $q=2$ and $\beta=1$, $q=12$). 
We have computed the risk-aware estimates empirically by using the primal representation of mean-upper-semideviation \eqref{my7} under the nominal noise model (that is, \textit{no} information about any possible alternative reality is needed). As anticipated, 
the risk-neutral estimate (expectation) 
masks the potential dispersion of the noise energy
(in fact, in every case),
with the issue being more
pronounced for the two alternative realities, and especially for the alternative reality whose noise distribution exhibits a fatter tail. 
In contrast, the risk-aware
estimates are 
biased toward 
more extreme noise energy realizations
(especially
in the case of $q=12$), which equivalently (via risk duality) provide
uniformly cautious estimates for a variety of possible 
realities,
including $w_{t,1}^{2}$ and $w_{t,2}^{2}$, 
\textcolor{black}{due to the richness of the uncertainty set \eqref{7}}.
These facts are readily apparent by observing  
the noise energy trajectories (along with the horizontal
lines) and the empirical CDFs (along with the vertical lines) in Figure \ref{Realities}. 
\end{illusexample}
\begin{illusexample}[Trade-offs related to $\kappa$]\label{illus2} Here, we will illustrate trade-offs related to the parameter $\kappa$ from Theorem \ref{th2}. Let us consider the case of $\eta < 1$, where the matrices $R$, $A$, and $H$ have been chosen. 
%
%
The parameters $\lambda$ and $c$ depend on an adjustable parameter $\kappa \in (0,1)$. Since $\lambda = 1 - \kappa \eta$ \eqref{myparamth2}, where $\eta \in (0,1)$ is determined by $R$, $A$, and $H$, choosing $\kappa$ to be closer to 1 would provide faster decay. However, we would also like $c$ to be small to reduce noise effects. 
(Recall that $b = c b'$ \eqref{myparamth2}, where $b'$ \eqref{mybprime} is the supremum of the risk of the noise energy $w_t^\top H_R w_t$.) There is a nonlinear relationship between $c$ and $\kappa$, that is, by direct substitution, we obtain
\begin{equation}
    c = \frac{1 - \kappa \eta}{\lambda_{\text{min}}(H) \; \kappa \eta \; (\eta - \kappa \eta)}.
\end{equation}
Figure \ref{cvskappa} depicts $c \lambda_{\text{min}}(H)$ versus $\kappa$ for several values of $\eta \in (0,1)$, illustrating the effect of $\eta$ on the minimum value of $c \lambda_{\text{min}}(H)$ with respect to $\kappa$. 
\begin{figure}
  \centering
  \includegraphics[scale = 0.6]{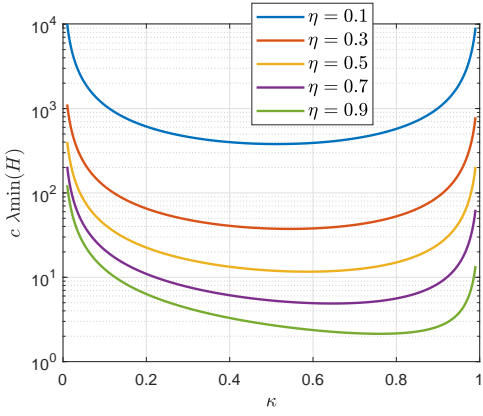}
  \caption{Plots of $c \lambda_{\text{min}}(H)$ versus $\kappa \in (0,1)$ on a semilog scale for several values of $\eta \in (0,1)$ (see Theorem \ref{th2} in the case of $\eta < 1$).}
  \label{cvskappa}
  \vspace{-10pt}
\end{figure}
\end{illusexample}

\begin{illusexample}[Insights about controllers using Theorem \ref{th3}]\label{illus3}
 An exciting potential use case 
 of the risk-aware stability
bounds we have developed 
is to inform the design of risk-aware controllers. The mean-conditional-variance functional studied in Theorem 3 turns out to be particularly useful for
this purpose, mainly due to its quadratic form. To demonstrate this,
suppose we are given a linear system of the form
\begin{equation}
x_{t+1}=\check{A}x_{t}+\check{B}u_{t}+\check{w}_{t},\quad t\in\mathbb{N}_{0},\label{eq:Controlled_System}
\end{equation}
where, for simplicity, we assume that the noise process $\check{w}_{t}$
is stationary but with possibly nontrivial dispersive behavior (to be described). For each $t\in\mathbb{N}_{0}$, we consider the \textit{myopic}
(one-step-ahead) regularized optimal control 
problem
\begin{equation}
\begin{array}{rl}
\underset{u_{t}\in\mathbb{R}^{m}}{\mathrm{minimize}} & \rho_{\nu}(\psi(x_{t+1})|x_{t})+\mu u_{t}^{\top}u_{t},\quad\mu\ge0,\end{array}
\end{equation}
where $\rho_{\nu}(\cdot|x_{t})$ denotes the conditional version
of $\rho_{\nu}$ (note
that $\rho_{\nu}$ admits an expectation representation, see the first line of \eqref{my5656}), when
the state $x_{t}$ at time $t$ is given. This problem is equivalent
to the quadratic program
\begin{equation}
\hspace{-1bp}\hspace{-1bp}\hspace{-1bp}\hspace{-1bp}\hspace{-1bp}\hspace{-1bp}\hspace{-1bp}\hspace{-1bp}\hspace{-1bp}\hspace{-1bp}\begin{array}{rl}
\underset{u_{t}\in\mathbb{R}^{m}}{\mathrm{minimize}} & \hspace{-1bp}\hspace{-1bp}\hspace{-1bp}\hspace{-1bp}\hspace{-1bp}\hspace{-1bp}E(x_{t+1}^{\top}R_{\nu}x_{t+1}\hspace{-1bp}+\hspace{-1bp}4\nu x_{t+1}^{\top}R\gamma|x_{t})\hspace{-1bp}+\hspace{-1bp}\mu u_{t}^{\top}u_{t},\end{array}\hspace{-1bp}\hspace{-1bp}\hspace{-1bp}\hspace{-1bp}\hspace{-1bp}\hspace{-1bp}
\end{equation}
where $\gamma_{t}=\gamma$ and $\Sigma_t = \Sigma_u$ for every $t$. Provided that $\mu I_n+\check{B}^{\top}R_{\nu}\check{B}$ is invertible, the solution
is
\begin{equation}
u_{t}^{\nu,\mu}=-K_{\nu}^{\mu}x_{t}-T_{\nu}^{\mu}\gamma,\quad t\in\mathbb{N}_{0},\label{eq:Controller_1}
\end{equation}
where the \textit{inflated gain} $K_{\nu}^{\mu}$ and \textit{tail
bias} $T_{\nu}^{\mu}$ are given by
\begin{align}
K_{\nu}^{\mu} & \coloneqq(\mu I_n+\check{B}^{\top}R_{\nu}\check{B})^{-1}\check{B}^{\top}R_{\nu}\check{A}\quad\text{and}\\
T_{\nu}^{\mu} & \coloneqq2\nu( \mu I_n +\check{B}^{\top}R_{\nu}\check{B})^{-1}\check{B}^{\top}R.
\end{align}
Observe that, when $\nu=0$, we obtain the standard myopic linear-quadratic-regulator controller $u_{t}^{0,\mu}=-K_{0}^{\mu}x_{t}$. Also, note
that the risk-aware controller $u_{t}^{\nu,\mu}$ is distinct from
the controller recently developed in \cite{tsiamis2020risk}, the latter being non-myopic
and derived via stochastic dynamic programming. 

\begin{figure}
  \centering
  \includegraphics[scale = 0.446]{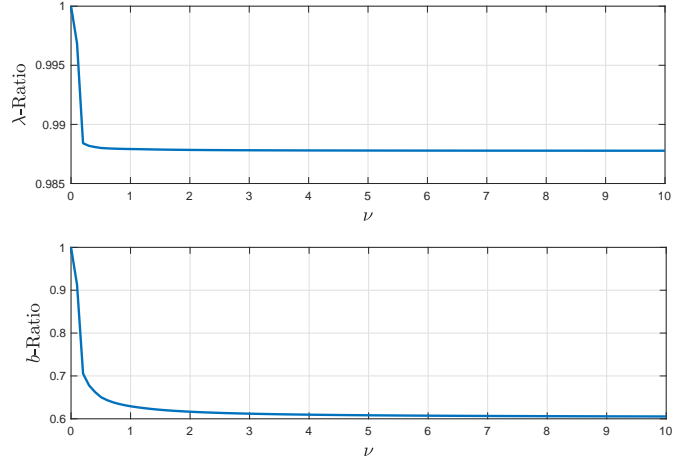}
  \caption{For different values of $\nu$, we show the rate-ratio (top) and bias-ratio (bottom) achieved by using the risk-aware controller relative to the risk-neutral controller for the stochastic linear system described in Illustration \ref{illus3}. ($\lambda$-ratio means rate-ratio. $b$-ratio means bias-ratio.) The bias-ratio plot indicates that the risk-aware controller exhibits a disturbance attenuation effect that becomes more pronounced as $\nu$ increases and saturates near $\nu = 10$.}
  \label{ratios}
  \vspace{-10pt}
\end{figure}

By substituting the stationary controller (\ref{eq:Controller_1}) into (\ref{eq:Controlled_System}),
we obtain the closed-loop system
\begin{align}
x_{t+1} & =(\check{A}-\check{B}K_{\nu}^{\mu})x_{t}+(\check{w}_{t}-\check{B}T_{\nu}^{\mu}\gamma)\nonumber \\
 & \eqqcolon Ax_{t}+w_{t},\quad t\in\mathbb{N}_{0},\label{eq:Controlled_System_2}
\end{align}
where, in contrast to the risk-neutral case ($\nu=0$), the risk-aware
controller not only regulates the state (in a risk-aware sense through
the inflated gain $K_{\nu}^{\mu}$, driving the state away from directions
with high second-order variability, as captured by $R_{\nu}$), but
also shifts the noise by a quantity proportional to its third-order
moment behavior, as captured by the statistic $\gamma$. Essentially,
the risk-aware controller optimally ``de-biases" the skewed behavior of $\check{w}_{t}$
by simply shifting its mean. This implies that the shifted noise $w_{t}$ has the same \textit{central} statistics (in particular, $\gamma$, $\delta$, and covariance) as the original noise $\check{w}_{t}$. 
%
%

Figure \ref{RealitiesController} shows the time trajectories and empirical CDFs of
the state energy and the respective control effort (where applicable)
corresponding to no controller ($u_t$ is zero), the risk-neutral controller
for $\mu=0.25$, and the risk-aware controller for $\mu=0.25$ and
$\nu=10$, for a Schur stable system of the form of (\ref{eq:Controlled_System})
by choosing
\begin{equation}
\check{A}=\begin{bmatrix}0.8 & 0.4\\
0 & -0.8
\end{bmatrix},\quad\check{B}=\begin{bmatrix}0\\
1
\end{bmatrix},\quad\text{and}\quad R=I_{2},
\end{equation}
and where $\check{w}_{t}$ follows a Gaussian mixture distribution 
\begin{equation}
\check{w}_{t}\overset{\text{iid}}{\sim}0.7\times{\cal N}(0_2,I_{2})+0.3\times{\cal N}([2\,\,15]^{\top},10I_{2}),\,\, t\in\mathbb{N}_{0}.
\end{equation}
In other words, for $70\%$ of the time, the additive disturbance to
the system is standard Gaussian, while for the remaining $30\%$ of
the time, the system exhibits abrupt Gaussian shocks with large mean
and variance in both state coordinates. 

We observe that, for the same level of control regularization ($\mu=0.25$), the risk-aware controller offers a dramatic improvement
on the state energy $|x_{t}|^{2}$ in terms of stabilizing its statistical
variability (i.e., risky behavior) as compared to its risk-neutral
counterpart, and it is particularly effective in mitigating the (more
infrequent) shocks due to the highly dispersive noise $\check{w}_{t}$.
Additionally, the control effort $|u_t|^2$ of the risk-aware controller is significantly
smaller \textit{in magnitude}, however less sparse and more persistent
as compared to the risk-neutral controller. This is explained due
to the strategically designed affine form of the risk-aware controller,
which provides increased degrees of freedom, allowing more effective
state regulation. 


Interestingly, the risk-aware controller $u_{t}^{\nu,\mu}$ exhibits
\textit{disturbance attenuation} behavior in the sense
of the risk functional $\rho_{\nu}$. Indeed,
such an effect can be readily observed through an empirical application
of Theorem 3, which also reveals the usefulness of Theorem 3 for informing and evaluating different controller designs.
%
%
To demonstrate this, in Figure \ref{ratios}
we report, for each value of $\nu$, the ratios of the biases $b_{\nu}$
(respectively, decay rates $\lambda_{\nu}$) achieved by using the
risk-aware controller $u_{t}^{\nu,\mu}$ over the risk-neutral controller
$u_{t}^{0,\mu}$. From Figure \ref{ratios} (bottom), we 
observe
that, as $\nu$ increases, the corresponding 
bias-ratio consolidates
sharply to a value roughly equal to $0.6$. In the context of Theorem
3, this finding implies a reduction in the bias term $b_\nu$ of our stability bound
as much as $40\%$ (roughly achieved for $\nu=10$) compared to
the bias achieved by the standard risk-neutral controller. Such a
reduction demonstrates a drastic disturbance attenuation effect exhibited by
the risk-aware controller $u_{t}^{\nu,\mu}$. 
%
Lastly, we observe a similar decreasing trend
for the corresponding rate-ratio (see Figure \ref{ratios}
(top)). Hence, the risk-aware controller also improves (albeit slightly)
the decay rate of the exponentially decreasing terms of the bound of
Theorem 3. 
\end{illusexample}
\section{Conclusion}\label{secVI}
By investigating a generalized risk-aware stability viewpoint, 
we have discovered conditions that guarantee new risk-aware noise-to-state stability properties for linear systems. In the case of any real-valued coherent risk functional on $\mathcal{L}^q$, the long-term risk of the state energy is on the same order as the supremum of the risk of the noise energy \eqref{242424}. In the case of a mean-conditional-variance functional, a similar relationship appears in \eqref{my4747}, and additionally, the noise-dependent bias term $b_{\nu}$ can be attenuated by a simple risk-aware controller (Illustration \ref{illus3}). 
%
%
In the future, we plan to investigate 
extensions to nonlinear systems, including systems with incomplete or misspecified models using statistical learning techniques.
%
%
We are particularly excited about extending the risk-aware stability theory developed here to enrich the analysis and design of stochastic gradient descent algorithms. 
 

%
\section*{Appendix}\label{secVII}
%
%
%
%
%
%
%
%
%
%

The following lemma 
can be viewed as a corollary of the classical discrete-time Lyapunov stability theorem for deterministic linear systems 
\cite[Th. 7.3.2]{datta2004numerical}.
\begin{lemma}\label{lemma2appendix}
Let $R \in \mathcal{S}_n^+$ and $A \in \mathbb{R}^{n \times n}$ be given.
The existence of an $H \in \mathcal{S}_n^+$ such that $H_R - A^\top H_R A \in \mathcal{S}_n^+$ is equivalent to $A$ being Schur stable.
%
%
\end{lemma}
\hspace{-3mm}\begin{proof}
We will show one direction. (The omitted direction uses $H_R \in \mathcal{S}_n^+$ and involves left-multiplying by $v^*$ and right-multiplying by $v$, where $v \in \mathbb{C}^n$ is an eigenvector of $A$.)
%
%
%
%
%
%
%
Now, $A$ is Schur stable if and only if, for any $Q \in \mathcal{S}_n^+$, there is a unique $X \in \mathcal{S}_n^+$ such that $X - A^\top X A = Q$ \cite[Th. 7.3.2]{datta2004numerical}. Assume that $A$ is Schur stable, and let $Q \in \mathcal{S}_n^+$ be given. Since $R^{\frac{1}{2}}$ is nonsingular, consider $H \coloneqq (R^{-\frac{1}{2}})^\top X R^{-\frac{1}{2}}$. Since $R^{-\frac{1}{2}}$ is nonsingular and $X \in \mathcal{S}_n^+$, $H \in \mathcal{S}_n^+$ holds as well. Moreover, since $H_R = X$, we have shown the existence of an $H \in \mathcal{S}_n^+$ such that $H_R - A^\top H_R A \in \mathcal{S}_n^+$.
\end{proof}

\begin{lemma}\label{etalemma}
Let $R \in \mathcal{S}_n^+$ and $A \in \mathbb{R}^{n \times n}$ be given.
Suppose that there is an $H \in \mathcal{S}_n^+$ such that 
$H_R - A^\top H_R A \in \mathcal{S}_n^+$. 
Then, $\eta \coloneqq \frac{\lambda_{\text{min}}(H_R - A^\top H_R A)}{\lambda_{\text{max}}(H_R)} \in (0,1]$.
\end{lemma}
\hspace{-4mm}\begin{proof}
The property $\eta \in (0,\infty)$ holds because $H_R - A^\top H_R A \in \mathcal{S}_n^+$ and $H_R \in \mathcal{S}_n^+$. If
\begin{equation}\label{30}
    \lambda_{\text{max}}(H_R - A^\top H_R A) \leq \lambda_{\text{max}}(H_R),
\end{equation}
then $\eta \leq 1$, so it suffices to show \eqref{30}. Since $A^\top H_R A \in \mathcal{S}_n$, the inequality \eqref{30} follows from the eigenvalue monotonicity theorem \cite[Cor. 4.3.12]{horn2012matrix}. 
\end{proof}

\begin{lemma}\label{keyu}
For any $M \in \mathcal{S}_n$, $y \in \mathbb{R}^n$, and $z \in \mathbb{R}^n$, and for every $\varepsilon \in (0,\infty)$, it is true that \vspace{-1mm} \begin{equation}\label{36}
\textstyle  (y + z)^\top M (y + z) \leq (1+\varepsilon) y^\top M y +  ( 1+ \frac{1}{\varepsilon} ) z^\top M z.
\end{equation}
\end{lemma}

Lemma \ref{keyu} is standard, and so we omit the proof. 

\begin{lemma}\label{quadformlemma}
Let $\mathcal{A}$ be the family of densities in the dual representation \eqref{dualrep} of a real-valued coherent risk functional $\varrho$ on $\mathcal{L}^q$ with $q \in [1,\infty)$. Consider the linear system \eqref{my88} described in the first paragraph of Section \ref{lin}, and assume that $|w_t|^2 \in \mathcal{L}^q$ for every $t \in \mathbb{N}_0$.
\begin{enumerate}
    \item For any $t \in \mathbb{N}_0$, $\xi \in \mathcal{A}$, and $M \in \mathcal{S}_n$, $x_t^\top M x_t \in \mathcal{L}^q$ and $x_t^\top M x_t \xi \in \mathcal{L}^1$, and these functions are a.e.-nonnegative.
    \item Given $H \in \mathcal{S}_n^+$ and $R \in \mathcal{S}_n^+$, define $\psi(z) \coloneqq z^\top R z$ and $v(z) \coloneqq z^\top H_R z$ for every $z \in \mathbb{R}^n$. For every $\xi \in \mathcal{A}$ and $t \in \mathbb{N}_0$, the statement \eqref{statement22} holds.
\end{enumerate}
 %
%
\end{lemma}
\hspace{-4mm}\begin{proof}
Part 1: $x_t^\top M x_t$ is a sum of finitely many $(\mathcal{F}, \mathcal{B}_{\mathbb{R}})$-measurable functions, and so it is $(\mathcal{F}, \mathcal{B}_{\mathbb{R}})$-measurable. The property of $|x_t|^2 \in \mathcal{L}^q$ for every $t \in \mathbb{N}_0$ holds by induction, where we use $|w_t|^2 \in \mathcal{L}^q$ and $x_{t+1} = A x_t + w_t$ with $x_0$ being fixed. 
In addition, we use
\begin{equation}
    0 \leq |x_{t+1}(\omega)|^2 \leq 2 \lambda_{\text{max}}(A^\top A) |x_t(\omega)|^2 + 2 |w_t(\omega)|^2
\end{equation}
for every $\omega \in \Omega$ (see Lemma \ref{keyu}) and $y \mapsto y^\gamma$ being nondecreasing on $[0,\infty)$ for any $\gamma \in (0,\infty)$ to find 
\begin{equation}
    \big \| \, |x_{t+1}|^2 \, \big \|_q \leq  \big \| \,2 \lambda_{\text{max}}(A^\top A) |x_t|^2 + 2 |w_t|^2 \, \big \|_q. 
\end{equation}
The property of $x_t^\top M x_t \in \mathcal{L}^q$ and $x_t^\top M x_t$ being nonnegative follow from
$0 \leq x_t^\top M x_t \leq \lambda_{\text{max}}(M) |x_t|^2$, $y \mapsto y^\gamma$ being nondecreasing on $[0,\infty)$ for any $\gamma \in (0,\infty)$, $|x_t|^2 \in \mathcal{L}^q$, and $\big \| \lambda_{\text{max}}(M) |x_t|^2 \big \|_q = \lambda_{\text{max}}(M) \big \| |x_t|^2 \big \|_q$.
%
%
%
%
%


Now, $\xi \in \mathcal{A}$ implies that $\xi \geq 0$ a.e. and $\xi \in \mathcal{L}^{q*}$. Hence, $x_t^\top M x_t \xi \geq 0$ a.e. and $\| x_t^\top M x_t \xi \|_1 \leq \|x_t^\top M x_t \|_q \, \| \xi \|_{q*} < \infty$ \cite[H\"{o}lder's Inequality, Th. 6.8 (a)]{folland1999real}. 
%
%

Part 2: Since $H \in \mathcal{S}_n^+$,
\begin{equation}\label{33}
    0 \leq \lambda_{\text{min}}(H)  |z|^2 \leq z^\top H z \leq  \lambda_{\text{max}}(H)  |z|^2, \quad z \in \mathbb{R}^n.
\end{equation}
Given $y \in \mathbb{R}^n$, consider $z = R^{\frac{1}{2}}y$ in \eqref{33} to find
\begin{equation} 
    0 \leq \lambda_{\text{min}}(H) y^\top R y \leq y^\top H_R y \leq  \lambda_{\text{max}}(H) y^\top R y,
\end{equation}
where we use $R = (R^{\frac{1}{2}})^\top R^{\frac{1}{2}}$ and $H_R = (R^{\frac{1}{2}})^\top H R^{\frac{1}{2}}$. We substitute $\psi(y) = y^\top R y$ and $v(y) = y^\top H_R y$ to find 
\begin{equation}\label{35}
    0 \leq \lambda_{\text{min}}(H) \psi(y) \leq v(y) \leq  \lambda_{\text{max}}(H)  \psi(y).
\end{equation}
Now, let $\xi \in \mathcal{A}$ and $t \in \mathbb{N}_0$ be given.
Since \eqref{35} holds for every $y \in \mathbb{R}^n$ and since $\xi \geq 0$ a.e.,
\begin{equation}\label{3687}
      0 \leq \lambda_{\text{min}}(H) \psi(x_t)\xi \leq v(x_t)\xi \leq  \lambda_{\text{max}}(H)  \psi(x_t)\xi \quad \text{a.e.}
\end{equation}
The statement \eqref{statement22} follows from \eqref{3687} and basic integration properties \cite[Th. 1.5.9, see also p. 47]{ash1972probability}.
%
%
\end{proof}
\begin{lemma}\label{geom}
Suppose that $\lambda \in (0,1)$, $\nu \in [0,\infty)$, and $(s_t)_{t \in \mathbb{N}_0}$ is a sequence in $[0,\infty)$ such that
   $  s_{t+1} \leq \lambda s_t + \nu $ for every  $t \in \mathbb{N}_0$.
Then,
   $ 0 \leq s_t \leq \lambda^t s_0 + \frac{\nu}{1-\lambda} $ for every $t \in \mathbb{N}$.
\end{lemma}
\hspace{-4mm}\begin{proof}
By induction, $0 \leq s_t \leq \lambda^t s_0 + \nu \sum_{i=0}^{t-1} \lambda^i$ for every $t \in \mathbb{N}$. Since $\lambda \in (0,1)$ and $\nu \in [0,\infty)$, we use the geometric series formula to find that $\nu\sum_{i=0}^{t-1} \lambda^i \leq \frac{\nu}{1-\lambda}$.
\end{proof}
%

%
\begin{lemma}\label{lemmakeyin}
Let $R \in \mathcal{S}_n^+$ and $A \in \mathbb{R}^{n \times n}$ be given. Suppose that there is an $H \in \mathcal{S}_n^+$ such that $H_R - A^\top H_R A \in \mathcal{S}_n$. Recall that $\eta \coloneqq \lambda_{\text{min}}(H_R - A^\top H_R A)/\lambda_{\text{max}}(H_R)$. Then, $\eta \in [0,1]$ and $  0 \leq z^\top A^\top H_R A z \leq (1-\eta)z^\top H_R z$ for every $z \in \mathbb{R}^n$.
\end{lemma}

\hspace{-8mm}\begin{proof}
Since $H_R - A^\top H_R A \in \mathcal{S}_n$ and $A^\top H_R A \in \mathcal{S}_n$, 
\begin{equation}\label{my3636}
    0 \leq \lambda_{\text{min}}(H_R - A^\top H_R A) \leq \lambda_{\text{max}}(H_R)
\end{equation}
by applying the eigenvalue monotonicity theorem \cite[Cor. 4.3.12]{horn2012matrix}. 
Since $H_R \in \mathcal{S}_n^+$, we divide \eqref{my3636} by $\lambda_{\text{max}}(H_R) \in (0,\infty)$ to find $0 \leq \eta \leq 1$. Now, let $z \in \mathbb{R}^n$ be given. We multiply the inequalities $0 \leq z^\top H_R z \leq \lambda_{\text{max}}(H_R) |z|^2$ by $\eta$ to find
\begin{equation}\label{my37}
    0 \leq \eta z^\top H_R z \leq \lambda_{\text{min}}(H_R - A^\top H_R A) |z|^2.
\end{equation}
We use \eqref{my37} and $H_R - A^\top H_R A \in \mathcal{S}_n$ to derive
\begin{equation}\label{my3737}
    \eta z^\top H_R z \leq z^\top (H_R - A^\top H_R A) z.
\end{equation}
Then, we use $A^\top H_R A \in \mathcal{S}_n$, and we add and subtract by $z^\top H_R z$ to find
\begin{align}
    0  \leq z^\top A^\top H_R A z & = -z^\top(H_R - A^\top H_R A) z + z^\top H_R z \nonumber \\
    & \leq -\eta z^\top H_R z + z^\top H_R z.
\end{align}
%
%
We have applied \eqref{my3737} in the last line to complete the proof.
\end{proof}

\begin{lemma}\label{show20}
Let the assumptions of Theorem \ref{th2} hold, and suppose that $\eta < 1$. Then, $ E(v(x_{t+1})\xi) \leq \lambda E(v(x_{t})\xi) + \frac{\lambda}{\lambda - (1-\eta)} b' $ for every $\xi \in \mathcal{A}$ and $t \in \mathbb{N}_0$.
\end{lemma}
\hspace{-3mm}\begin{proof}
Let $\xi \in \mathcal{A}$ and $t \in \mathbb{N}_0$ be given, and recall that $\xi \geq 0$ almost everywhere. We use $v(x_{t+1}) = x_{t+1}^\top H_R x_{t+1}$, $H_R \in \mathcal{S}_n^+$, $x_{t+1} = A x_t + w_t$, and Lemma \ref{keyu} to derive   
%
%
\begin{equation}\label{37}
    v(x_{t+1})\leq (1+\varepsilon) x_t^\top A^\top H_R A x_t  + \bigg( 1+ \frac{1}{\varepsilon}\bigg) w_t^\top H_R w_t
\end{equation}
for any $\varepsilon \in (0,\infty)$. Multiplying \eqref{37} by $\xi$, taking expectations, and using that $E(w_t^\top H_R w_t \xi) \leq b'$ from  \eqref{mybprime}
lead to
\begin{equation}\label{38}
    E(v(x_{t+1})\xi) \leq (1+\varepsilon)E(x_t^\top A^\top H_R A x_t \xi) + \bigg( 1+ \frac{1}{\varepsilon}\bigg)b'.
\end{equation}
Lemma \ref{keyu} helps circumvent the issue of the cross term $E(x_t^\top A^\top H_R w_t \xi)$ need not being zero. Due to \eqref{38} and $\lambda = 1-\kappa\eta$ \eqref{myparamth2} for any fixed $\kappa\in(0,1)$, it suffices to show that
 $   E(x_t^\top A^\top H_R A x_t \xi) \leq (1-\eta)E(v(x_t)\xi) $,
which readily follows in particular from Lemma \ref{lemmakeyin} and $\xi$ being nonnegative almost everywhere. To see this, note that
\begin{equation}
    (1+\varepsilon)(1-\eta)<1 \iff \varepsilon < \frac{\eta}{1-\eta}.
\end{equation}
By choosing $\varepsilon=(1-\kappa)\frac{\eta}{1-\eta}$, we obtain 
\begin{equation}
    \lambda = \bigg(1+(1-\kappa)\frac{\eta}{1-\eta}\bigg)(1-\eta)= 1-\kappa\eta.
\end{equation} 
Using the same value for $\varepsilon$ produces the quantity multiplying $b'$ in the statement of Lemma \ref{show20}.
\end{proof}

\begin{lemma}\label{bigquadform}
Consider
the linear system \eqref{my88}, and let $R\in{\cal S}_{n}^{+}$ and 
$\nu \in [0,\infty)$ 
be given. Consider the quadratic state energy function $\psi(x)\coloneqq x^{\top}Rx$. Under the assumptions of Theorem \ref{th3}, for every $t \in \mathbb{N}$, the following equality holds
\begin{equation}
    \rho_{\nu}(\psi(x_{t})) = E(x_{t}^{\top}(R+4\nu R\Sigma_{t}R)x_{t}+4\nu x_{t}^{\top}R\gamma_{t})+\mathbb{C}_{t},
\end{equation}
where $\mathbb{C}_{t} \coloneqq \nu\delta_{t}-4\nu\text{tr}((\Sigma_{t}R)^{2})$, and $\rho_{\nu}(\psi(x_{t}))$ is finite.
\end{lemma}
%
\hspace{-3mm}\begin{proof}
The proof is inspired by the proof of \cite[Prop. 1]{tsiamis2020risk} with additional steps to circumvent expressions of the form $\infty-\infty$.
Let $t \in \mathbb{N}$ be given. 
The assumptions of Theorem \ref{th3}
imply that $E(w_{t-1}) = \bar{w}_{t-1} \in \mathbb{R}^n$, $E(d_t d_t^\top) = \Sigma_t \in \mathcal{S}_n$, $\gamma_t \in \mathbb{R}^n$, and $\delta_t \in [0,\infty)$. 
(We use standard measure-theoretic principles, including H\"{o}lder's inequality \cite[Th. 2.4.5]{ash1972probability} and Minkowski's inequality \cite[Th. 2.4.7]{ash1972probability}. We do not use $\sup_{t\in\mathbb{N}_{0}}|\bar{{\bf w}}_{t}|^{2} < \infty$, $\sup_{t \in \mathbb{N}} |\gamma_t|^2 < \infty$, or $\sup_{t \in \mathbb{N}} \delta_t < \infty$.)
%
The conditional expectation $\hat{\chi}_t \coloneqq E(x_t|\mathcal{F}_{t-1})$ takes values in $\bar{\mathbb{R}}^n$ and is $(\mathcal{F}_{t-1}, \mathcal{B}_{\bar{\mathbb{R}}^n})$-measurable \cite[Th. 6.4.3, Th. 1.5.8]{ash1972probability}. 
%
%
One should be cautious about using $\hat{\chi}_t^\top R \hat{\chi}_t$ because it may take the ill-defined form of $\infty -\infty$. 
Hence, we will alter $\hat{\chi}_t$ on a set of measure one. Since $E(x_t) \in \mathbb{R}^n$, $\hat{\chi}_t$ is $\mathbb{R}^n$-valued almost everywhere \cite[Th. 6.5.4 (a), Th. 1.6.6 (a)]{ash1972probability}, 
and thus, the set 
 $   B_t \coloneqq \{\omega \in \Omega : \hat{\chi}_t(\omega) \in \mathbb{R}^n\} $
satisfies $P(B_t) = 1$. Moreover, $B_t \in \mathcal{F}_{t-1}$ because $\hat{\chi}_t$ is $(\mathcal{F}_{t-1},\mathcal{B}_{\bar{\mathbb{R}}^n})$-measurable.
%
%
We define $\hat{z}_t \coloneqq \mathcal{I}_{B_t} \hat{\chi}_t$. Hence, $\hat{z}_t$ is $(\mathcal{F}_{t-1},\mathcal{B}_{\bar{\mathbb{R}}^n})$-measurable, $\hat{z}_t(\omega) \in \mathbb{R}^n$ for every $\omega \in \Omega$, and $\hat{z}_t = \hat{\chi}_t$ almost everywhere. We use properties of conditional expectations \cite[Ch. 6.5]{ash1972probability}, $(\mathcal{F}_{t-1},\mathcal{B}_{\mathbb{R}^n})$-measurability of $x_{t-1}$, $x_t = A x_{t-1} + w_{t-1}$, and independence of $w_{t-1}$ and $h_{t-1}$ to find that $\hat{z}_t = x_t - d_t$ almost everywhere.
%
%
Moreover, we derive that $E(x_t^\top R x_t |\mathcal{F}_{t-1}) = \hat{z}_t^\top R \hat{z}_t + \text{tr}(\Sigma_t R)$ almost everywhere. Next, we apply the previous two statements, the definition of $\Delta_t$ \eqref{Deltat} with $Z = x_t^\top R x_t$ and $\mathcal{F}_i = \mathcal{F}_{t-1}$, and $\hat{z}_t^\top R \hat{z}_t$ being finite
to find that $\Delta_t = \tilde{\Delta}_t$ a.e., where $\tilde{\Delta}_t$ is defined by
\begin{equation}
    \tilde{\Delta}_t \coloneqq \text{tr}(\Sigma_t R) - 2 \hat{z}_t^\top R d_t -d_t^\top R d_t.
\end{equation}
%
%
%
In addition, all terms of $\tilde{\Delta}_t^2$ are integrable. (In particular, note that $\hat{z}_t$ and $d_t$ are independent and $E(d_t) = 0_n$.)
Then, we use $E(\tilde{\Delta}_t^2) = E(\Delta_t^2)$ and $\hat{z}_t = x_t - d_t$ a.e. to find 
\begin{align}
    E(\Delta_t^2 ) \hspace{-.5mm} = \hspace{-.5mm}
     4 E( x_t^\top R \Sigma_t R x_t) + 4 E( x_t^\top R \gamma_t) + \delta_t - 4 \text{tr}((\Sigma_t R)^2) , \label{61}
\end{align}
which is finite. Finally, using $\rho_\nu(x_t^\top R x_t) = E(x_t^\top R x_t) +\nu E(\Delta_t^2)$ completes the proof.
\end{proof}
%


\section*{Acknowledgement}
M.P.C. would like to gratefully acknowledge Erick Mejia Uzeda for discussions.

\bibliographystyle{IEEEtran}
\bibliography{references}

%
%


\end{document}